\documentclass[twocolumn]{aastex63}

\usepackage{amsmath,rotating} 
\renewcommand{\tablename}{Algorithm} 
\NewPageAfterKeywords

\vspace{-5mm}
\received{January 27, 2022}
\revised{March 16, 2022}
\accepted{April 9, 2022}
\submitjournal{ApJ}
\vspace{-15mm}
\shorttitle{Constraining Coronal Models}
\shortauthors{Badman et al.}
\graphicspath{{./}{figures/}}

\begin{document}

\title{Constraining Global Coronal Models with Multiple Independent Observables}

\correspondingauthor{Samuel T. Badman}
\email{samuel\_badman@berkeley.edu}

\author[0000-0002-6145-436X]{Samuel T. Badman}
\affil{Physics Department, University of California, Berkeley, CA 94720-7300, USA}
\affil{Space Sciences Laboratory, University of California, Berkeley, CA 94720-7450, USA}

\author[0000-0002-2189-9313]{David H. Brooks}
\affil{College of Science, George Mason University, 4400 University Drive, Fairfax, VA 22030 USA}
\affil{Hinode Team, ISAS/JAXA, 3-1-1 Yoshinodai, Chuo-ku, Sagamihara, Kanagawa 252-5210, Japan}

\author[0000-0002-1814-4673]{Nicolas Poirier}
\affil{IRAP, Université Toulouse III—Paul Sabatier, CNRS, CNES, Toulouse, France}

\author[0000-0001-6102-6851]{Harry P. Warren}
\affil{Space Science Division, Naval Research Laboratory, Washington, DC 20375, USA}

\author[0000-0001-8462-9161]{Gordon Petrie}
\affil{National Solar Observatory, 3665 Discovery Drive, 3rd Floor, Boulder, CO 80303, USA}

\author[0000-0003-4039-5767]{Alexis P. Rouillard}
\affil{IRAP, Université Toulouse III—Paul Sabatier, CNRS, CNES, Toulouse, France}

\author[0000-0001-9326-3448]{C. Nick Arge}
\affil{GSFC, Greenbelt, MD 20771}

\author[0000-0002-1989-3596]{Stuart D. Bale}
\affil{Physics Department, University of California, Berkeley, CA 94720-7300, USA}
\affil{Space Sciences Laboratory, University of California, Berkeley, CA 94720-7450, USA}
\affil{The Blackett Laboratory, Imperial College London, London, SW7 2AZ, UK}

\author[0000-0002-5469-2378]{Diego de Pablos Ag\"uero}
\affil{UCL-MSSL, Holmbury St. Mary, Dorking, Surrey RH5 6NT, UK}

\author[0000-0001-9457-6200]{Louise Harra}
\affil{PMOD/WRC, Dorfstrasse 33 CH-7260 Davos Dorf, Switzerland}
\affil{ETH-Zurich, H\"onggerberg campus, HIT building, Z\"urich, Switzerland}

\author[0000-0001-9498-460X]{Shaela I. Jones}
\affil{GSFC, Greenbelt, MD 20771}

\author[0000-0001-6589-4509]{Athanasios Kouloumvakos}
\affil{IRAP, Université Toulouse III—Paul Sabatier, CNRS, CNES, Toulouse, France}

\author[0000-0002-1859-456X]{Pete Riley}
\affil{Predictive Sciences Inc., San Diego, CA 92121}

\author[0000-0002-4440-7166]{Olga Panasenco}
\affil{Advanced Heliophysics, Pasadena, CA 91106, USA}

\author[0000-0002-2381-3106]{Marco Velli}
\affil{EPSS, UCLA, Los Angeles, CA 90095, USA}

\author[0000-0002-1091-4688]{Samantha Wallace}
\affil{GSFC, Greenbelt, MD 20771}



\begin{abstract}

Global coronal models seek to produce an accurate physical representation of the Sun's atmosphere which can be used, for example, to drive space weather models. Assessing their accuracy is a complex task and there are multiple observational pathways to provide constraints and tune model parameters. Here, we combine several such independent constraints, defining a model-agnostic framework for standardized comparison. We require models to predict the distribution of coronal holes at the photosphere, and neutral line topology at the model outer boundary. We compare these predictions to extreme ultraviolet (EUV) observations of coronal hole locations,  white-light Carrington maps of the streamer belt and the magnetic sector structure measured \textit{in situ} by Parker Solar Probe and 1AU spacecraft. We study these metrics for Potential Field Source Surface (PFSS) models as a function of source surface height and magnetogram choice, as well as comparing to the more physical Wang-Sheeley-Arge (WSA) and the Magnetohydrodynamics Algorithm outside a Sphere (MAS) models. We find that simultaneous optimization of PFSS models to all three metrics is not currently possible, implying a trade-off between the quality of representation of coronal holes and streamer belt topology. WSA and MAS results show the additional physics they include addresses this by flattening the streamer belt while maintaining coronal hole sizes, with MAS also improving coronal hole representation relative to WSA. We conclude that this framework is highly useful for  inter- and intra-model comparisons. Integral to the framework is the standardization of observables required of each model, evaluating different model aspects. 

\end{abstract}
\vspace{-5mm}
\keywords{Sun:corona -- Sun:magnetic fields -- Sun: evolution}


\section{Introduction} \label{sec:intro}


The solar corona is the tenuous outer layer of the Sun's atmosphere extending out to a few solar radii which comprises hot magnetized plasma.  This system is of critical importance for understanding how the Sun interacts with its local environment by creating the solar wind, and how it can affect the Earth. While much can be learned about the corona indirectly from remote observations including plasma density and temperature structure \citep[e.g.][]{Warren2009}  or magnetic topology \citep[e.g.][]{Boe2020}, such investigations are usually limited to the plane of the sky as viewed from Earth or one of the STEREO spacecraft. With direct \textit{in situ} measurements within 1-5$R_\odot$ currently technologically out of reach, a comprehensive synchronic representation of the full 3D structure can only currently be achieved using physics-based modeling.

Accurately reproducing this structure with models is of vital importance for innumerable endeavours in solar and heliophysics. For example, coronal models form the central component of space weather simulation and prediction frameworks \citep[see e.g. review by][]{Feng2020}, they enable contextualization and prediction of \textit{in situ} spacecraft data \citep[e.g. ][]{Riley2019,Badman2020} and ultimately hold the key to fundamental plasma physics problems that are still unsolved, including coronal heating.

The most commonly used, and one of the earliest implemented coronal models was the Potential Field Source Surface (PFSS) model \citep{Altschuler1969,Schatten1969} wherein the corona is assumed to be magnetostatic (i.e. current-free) in an annular domain with an inner boundary of radial magnetic field derived from direct magnetograph observations of the photosphere, and the outer boundary assumed to be an equipotential of radius $R_{SS}$, called the source surface. (Typically this is set to a value around $2.5 R_\odot$ \citep[][]{Hoeksema1984}). The equipotential surface simulates the outflowing solar wind by forcing field lines intersecting it to be radial and produces ``open'' field lines along which plasma can escape, without needing detailed plasma modeling. 

PFSS models have some obvious drawbacks. While they produce open field typically rooted in observed coronal holes, the highest altitude closed loops are rounded while eclipse observations show that in fact the last closed loops are kinked, and spike outwards in a streamer configuration \citep{Altschuler1969}. Further, the height where field lines are observed to become uniformly radial differs not just from the standard \citep{Hoeksema1984} $2.5R_\odot$ value, but also varies from location to location on the Sun \citep{Riley2006,Boe2020} and in time with the evolution of the solar cycle \citep{Reville2017}. PFSS models also only directly predict coronal magnetic structure, and have an indirect connection to solar wind flow speed \citep[][]{Wang1992} but otherwise pose no constraints on other plasma properties such as density, temperature or composition. They are also manifestly time-independent. As such, numerous more realistic models have been developed, usually trading improved physical content at the price of computational tractability.

PFSS extrapolations can be extended and improved beyond $2.5$R$_\odot$ with a Schatten Current Sheet \citep{Schatten1972}, as implemented in the Wang-Sheeley-Arge (WSA) model \cite{Arge2000,Arge2003,Arge2004}, which produces more kink-shaped streamer topology and a semi-empirical solar wind velocity field which is a function of field line expansion factor and the distance between traced open fieldline footpoints and coronal hole boundaries.

Magnetohydrodynamic (MHD) simulations of the solar corona represent the state of the art of global coronal modeling. To the extent that MHD represents the true physical processes in the Sun's atmosphere, these models constitute a complete, time dependent representation of the associated plasma flows including plasma moments, wave normal modes and instabilities. Variations between MHD models of the corona often stem from how energy is supplied to the corona and hence drives supersonic outflow, and generally come down to the energy equation in the underlying numerical scheme. Numerous model implementations for the Sun's corona exist; examples include the Alfven Wave Solar Model \citep[AWSoM; ][]{Sokolov2013,vanderHolst2014}, the Multi-Scale Fluid-Kinetic Simulation Suite \citep[MS-FLUKSS; ][]{Pogorelov2013} and the Magnetohydrodynamics Around a Sphere \citep[e.g. ][]{Mikic1996,Mikic1999}, the latter of which is used in the present work.

Although the three categories of models mentioned above comprise those examined in this article, there are numerous other coronal models that are being developed and seek to improve on PFSS models without going as far as full MHD. For example \citet{Kruse2020} has produced a PFSS extension with an elliptical source surface. \citet{Rice2021} proposed a magnetofrictional model which is similar to PFSS models except it produces its coronal field self consistently with an accelerating radially outwards flow which leads to much more compelling streamer belt shapes. Nonlinear Force Free (NLFF) models \citep[see review by ][]{Wiegelmann2021}  are also an improvement on potential fields in that they reduce the severity of the no-current constraint, allowing currents to flow along field lines.

It is clear that there are a plethora of options for coronal modeling, with trade-offs between the physical content of the model, computational time and accessibility of model software. However, there is an implicit assumption here that ``more physics'' should produce more accurate coronal models. To support this assumption, it is clearly desirable to compare models to each other and assess for improvements in accuracy.

Almost any publication introducing a new or tuned coronal model includes a qualitative or quantitative comparison to empirical measurements of the Sun (either remote or in situ). For example, \citet{Lee2011}, \citet{Arden2014}, \citet{Badman2020} and \citet{Panasenco2020} all investigated PFSS source surface height with the first comparing the extent of coronal holes, the second trying to match the open magnetic flux in the heliosphere and the latter two trying to match magnetic polarity changes with Parker Solar Probe measurements. Similarly, \citet{Kruse2021} used ballistically mapped magnetic polarity to study an elliptical outer boundary condition to a potential model. Even the first PFSS papers investigated different observables where \citet{Altschuler1969} focused on comparison to white light (WL) eclipse observations while \citet{Schatten1969} investigated interplanetary magnetic field strength. While this single metric approach is usually useful to support an author's assertion of improvement, especially when comparing their model to a prior iteration of the same model, there are a couple of issues. Firstly, such comparisons made with one type of metric (for example prediction of current sheet crossing timing, or visual correspondence of model field lines to a coronagraph snapshot) will test and optimize a single aspect of the models' representation of the corona. As we show in this work (section \ref{subsec:results1-vs_rss}), it is possible to improve one aspect of a model to the detriment of another. Secondly, this leads to highly non-standardised comparisons across the literature. For illustration, we may know model X produces a more accurate forward model of an eclipse observation than model Y, and that model Y produces a much more accurate timeseries  prediction of velocities compared to the \textit{in situ} measurements than model Z. However this does not allow us to compare model X to model Z, without significant further work. 

The goal of this work is to define a model-agnostic framework to evaluate quantitatively the representation of the corona for different coronal models and to use multiple independent comparisons to do so. We study PFSS, WSA and MAS MHD model results and for consistency, we study the minimum set of information which is provided directly by all three models, i.e. the coronal magnetic field. For each model, we require a prediction in a standardized format of 1) the location of coronal holes at the coronal base (i.e. the location of open magnetic field in the model), and 2) the location of the coronal neutral line at the outer boundary of each model which corresponds to the coronal streamer belt and shapes the heliospheric current sheet (HCS). To test these predictions, we identify three independent observations. The first is EUV observations of the locations of coronal holes, the second is the location of the streamer belt at $5R_S$ observed from white light coronagraphs, and last is the timing of crossings of the heliospheric current sheet as measured by heliospheric spacecraft (Parker Solar Probe, ``near-earth'' spacecraft via the OMNI dataset, and STEREO A). Each of these three types of observation is compared to the relevant model prediction and a quantitative metric is developed to compare them. We study time intervals of $\pm$30 days from the date of the first three perihelia of Parker Solar Probe (PSP) to focus our investigation and take advantage of novel \textit{in situ} measurements of PSP closer to the Sun than ever before. Our metric scores are probed as a function of the choice of model, input parameters, boundary conditions and time over our model dataset. We present results and inferences for these specific studies (Section \ref{sec:results}) as well as more general conclusions about our model framework (Section \ref{sec:discussion}).

Similar issues have recently been discussed and studied in complementary recent work by \citet{Wagner2021} with the implementation goal of optimizing the EUropean Heliospheric FORecasting Information Asset \citep[EUHFORIA;][]{Pomoell2018} model, another example of an MHD model. As in the present work, these authors specify individual independent evaluation metrics between models and observations. In this work, where relevant, we explain the relationship or difference of implementation in our metrics as compared to those of \citet{Wagner2021}. 

\section{Coronal Field Models} \label{sec:models}

As mentioned above, several coronal models are used in this work to define and test our evaluation framework. In this section, we introduce the model types and specific implementations used to then discuss the model-independent format of ``observables'' that each model is required to produce to facilitate quantitative comparisons.

Models were obtained for three specific time intervals, distributed $\pm$30 days either side of the first three perihelia of Parker Solar Probe (PSP). As mentioned in the introduction, this choice was made to focus our investigation and make use of novel new \textit{in situ} measurements from PSP. For each model, at least one run per interval was generated, but where possible provided full daily updates to the magnetogram input boundary conditions spanning these time intervals. 

\subsection{Potential Field Source Surface (PFSS) Models}\label{subsec:PFSS}

PFSS models for this work were generated from two sources: 

\begin{enumerate}
    \item  \textit{pfsspy}\footnote{\url{https://github.com/dstansby/pfsspy}}, an open source python implementation \citep{Stansby2020} with Air force Data Assimilative Flux Transport \citep[ADAPT; ][]{Arge2010,Arge2011,Arge2013,Hickmann2015} magnetograms as their input boundary conditions. ADAPT maps enhance typical magnetogram observations by assimilating direct measurements of magnetic fields on the solar near side and helioseismological sounding of far side structure into a self-consistent surface flux transport model. ADAPT map data products include 12 individual realizations obtained by running an ensemble of models. For the PFSS model run here, we average these realizations together, subtracting any residual monopole moment induced by the averaging before feeding it into \textit{pfsspy}. ADAPT maps are regularly produced using magnetographs from both Global Oscillation Network Group \citep[GONG][]{Harvey1996}  and from the Helioseismic and Magnetic Imager \citep[HMI; ][]{Scherrer2012} on board the Solar Dynamics Observatory \citep[SDO; ][]{Pesnell2012}. Both are used and their separate models generated with \textit{pfsspy} in this model dataset. In this work, these two magnetograms will be referred to as ``ADAPT-GONG'' and ``ADAPT-HMI'' respectively.
    \item The standard PFSS implementation by the Global Oscillation Network Group \citep[GONG; ][]{Harvey1996} which utilizes the zero-corrected GONG magnetogram data product \citep[][]{Hill2018}. This magnetogram will be referred to as ``GONGz'' in this work.
\end{enumerate}

As well as being tabulated for each day in the studied time intervals, individual model runs were produced for differing values of source surface height ($R_{SS}$) ranging from $1.5$ to $3.0R_\odot$, spanning most of the range typically investigated in studies of $R_{SS}$ \citep[e.g.][]{Lee2011,Arden2014}.

\subsection{Wang-Sheeley-Arge (WSA) Model}

The Wang-Sheeley-Arge (WSA) model \citep{Arge2000,Arge2003,Arge2004} extends the capabilities of PFSS models with a Schatten Current Sheet \citep[SCS][]{Schatten1972} model extending above the potential layer, providing magnetic coronal structure out to 5$R_\odot$. WSA additionally generates an empirical velocity field at the model outer boundary (generated as a function of expansion factor and the distance of magnetic footpoints to coronal holes) although only the magnetic structure is accessed in the present work. As with \textit{pfsspy}, for this project these models are driven by ADAPT magnetograms. Selection of magnetogram realization varied across different time intervals.

WSA model parameters include a source surface height ($R_{SS}$) marking the boundary between the PFSS and SCS components of the model, and an outer boundary for the SCS ($R_{SCS}$). Unless otherwise noted, these values were set to $R_{SS}=2.5R_\odot$ and $R_{SCS}=5R_\odot$.

Full daily evolved models were provided for our data set for the perihelia 1 and 3 intervals. For perihelion 2, a single model run was investigated using an ADAPT-GONG map from 2021-04-10. As will be discussed in section \ref{sec:results}, encounter 2 was challenging for modelers due to far side active region emergence. This single model run was chosen for encounter 2 as it provided the best representation of PSP in situ magnetic polarity.

\subsection{Magnetohydrodynamics Around a Sphere (MAS) MHD Model}

Lastly, we use results from the Magnetohydrodynamics Around a Sphere (MAS) model \citep{Mikic1994,Mikic1999,Riley2001} for the studied time intervals. As touched on in the introduction, MHD models are significantly more computationally intensive and as such daily evolution of the magnetogram input was not available for this project. Instead a single model run is provided for each perihelion date based on the closest Carrington rotation number (CRN).  

The MAS models extend from the corona out to $30R_\odot$. The supersonic solar wind outflow in this case is generated via the PSI thermodynamic energy equation \citep{Lionello2001,Lionello2009}. The magnetograms used to drive them from the inner boundary used HMI data \citep{Mikic2018b,Riley2019a}. Again, although the MAS model provides a litany of other physical properties, (e.g. density, flow velocity, temperature), only magnetic structure is extracted for this work.

\subsection{Derived Observables from Coronal Field Models}




A very important step in the framework outlined in this paper is defining a standardized set of ``observables'' to be produced by each model. This allows inter-model comparison and portability of this method to future models by ensuring that the comparison between observations and models is made in a controlled way that does not depend on any internal properties of the model used to produce them.

Specifically, we require models to produce:

\begin{itemize}
    \item A 2D open field map at the photosphere uniformly binned in sine heliographic latitude and heliographic Carrington longitude. Each pixel takes either the value 1 to indicate that a field line traced from that point on the photosphere reaches the model outer boundary and is magnetically ``open'', or 0 to indicate a field line traced from the point is ``closed'' and re-intersects the photosphere. The left hand edge of the 2D map is required to be at 0 degrees Carrington longitude.
    \item A 2D map of the position of the heliospheric current sheet at the outer boundary of the model binned in the same way as the coronal hole map. Pixels have the value of +1 for locations where the field is directed outwards (positive polarity) and -1 where the field is directed inwards (negative polarity). Presented in this way, the HCS can be extracted from the map using contouring software in IDL or python at the zero contour level, which allows for the possibility that the HCS consists of multiple closed loops. Although such topologies were not encountered in this study due to the deep solar minimum, this allows this framework to be useful at future (or previous) solar maxima, where such a topology may occur.
\end{itemize}

Thus, we require models to predict coronal structure at both their inner and outer boundaries. Examples of these two observables may be seen in the top panels of figures \ref{fig:metric1-illustration} and \ref{fig:metric2-illustration} respectively. Note we are only examining magnetic field structure. While coronal plasma of course comprises many other physical quantities (such as density, flow, temperature and composition), here we opt to restrict to magnetism since this is the minimal physical content common to all realistic coronal models and so allows direct comparison between all commonly used models, including those to come in the future.

\section{Observations and Measurements}\label{sec:obs}

In this section, we describe the corresponding observational data used and processed to facilitate comparison to the models. Data is obtained concurrently with the models, i.e. for the time intervals around each of PSP's first three perihelia.

\subsection{EUV Coronal Hole Synoptic Maps}
\label{ch_maps}
%

We use SDO Atmospheric Imaging Assembly \citep[AIA,][]{Lemen2012} observations to prepare our EUV Carrington maps from which coronal holes can be detected. We downloaded 180\,s worth of data for the 304\,\AA, 171\,\AA, 193\,\AA, and 211\,\AA\, filters each day of the Carrington rotations relevant to PSP encounters 1--3. The data were processed using standard AIA analysis software. We performed a deconvolution of the point-spread-function (PSF) on each image using \verb+deconvolve+ from aiapy\footnote{\url{https://aiapy.readthedocs.io/en/latest/}} and produced a median filtered composite image in each wavelength for each day. The daily images were converted from helioprojective coordinates to Carrington longitude and sine latitude using \verb+wcs_convert_to_coord+\footnote{\url{https://hesperia.gsfc.nasa.gov/ssw/gen/idl/wcs/conversion/wcs_convert_to_coord.pro}}, a routine in the Solar Software (SSW) IDL package  \citep{Freeland1998}. Finally, these are combined to produce an EUV Carrington map for the Carrington rotation. The blending overlap is about 12$^o$ in longitude. PSP perihelia 1 and 3 overlapped mostly with Carrington rotations 2210 and 2221 respectively, but encounter 2 occurred near the boundary of Carrington rotations 2215 and 2216. In this case, we prepared maps for both rotations, switching between them at the appropriate time.

Several methods of coronal hole detection have been used by the community dating back to early efforts using hand-drawn maps \citep{Harvey2002}. Subsequently, a number of independent detection methods have been developed. For example, using local intensity thresholding \citep{Krista2009}, forming high contrast images from linear combinations of spectral line properties \citep{Malanushenko2005}, segmenting images by region growing algorithms \citep{Caplan2016}, and applying intensity thresholding across multi-temperature image passbands \citep{Garton2018}. Each of these techniques has its advantages and weaknesses. Automatic detection algorithms, for example, are very powerful for processing large volumes of data. 

In this work we experimented with several methods: simple EUV intensity thresholding, the CHIMERA \citep[coronal hole identification via multi-thermal emission recognition algorithm,][]{Garton2018} software package, and the Predictive Science Inc.\ (PSI) EZSEG image segmentation code \citep{Caplan2016}. Ultimately we chose to use the EZSEG algorithm. For this work, the chosen algorithm needed to be able to accurately produce contours of coronal holes from 2D Carrington maps to allow direct comparison to our chosen observables (see section \ref{subsec:metric1}). By visual comparison, the EZSEG algorithm was found to be the best for our circumstances, particularly proving useful at contouring the polar coronal holes which form a large fraction of the global coronal hole area at solar minimum. 

The software is available as part of the EUV2CHM Matlab package from the Predictive Science Inc. website\footnote{\url{http://www.predsci.com/chd/}}. We converted the EZSEG algorithm from FORTRAN to IDL to plug-in to our software pipeline. This algorithm uses an initial intensity threshold to acquire coronal hole locations in an EUV image, and then uses an area growing technique to define connected regions. This continues until a second intensity threshold is reached, or the condition for connectivity is not met. The dual-thresholds and connectivity conditions (essentially the number of consecutive pixels), are defined on input. We experimented with the optimal input parameters for each of our 193\,\AA\, Carrington maps. As an example, for encounters 1 and 3, thresholds of $\sim$20\% and $\sim$60--70\% of the peak EUV intensity worked well, with connectivity defined as 5 consecutive pixels. For encounter two, thresholds of $\sim$15\% and $\sim$45\% worked better. This reflects the presence of several brighter active regions around the solar surface during encounter two. An example of our 193\,\AA\, Carrington maps and the EZSEG output is shown in the middle panel of figure \ref{fig:metric1-illustration}, and shows how the algorithm performs well at contouring coronal hole regions identified by eye as dark regions in the EUV carrington map. This figure as a whole and section \ref{subsec:metric1} explain how we make the comparison of our maps with modeled coronal hole areas.

\subsection{White Light Synoptic Maps}
\label{subsec:WL_Maps_subsection}


We exploit continuous white-light observations from the \textit{Solar and Heliospheric Observatory} \citep[\textit{SOHO}:][]{Domingo1995} LASCO C2 coronagraph \citep{Brueckner1995} to produce white-light synoptic maps. 

The light received by such a detector located at 1 AU is the result of multiple contributions along the lines-of-sight. The most significant contribution is the photospheric light scattered by dust particles, the F-corona. The latter is removed by subtracting background images from raw images in order to reveal the faint amount of photospheric light that is scattered by electrons, the K-corona. Hence this process provides not only access to the overall structure of the corona, but also to detail often observed in eclipse images for instance \citep{Boe2020,Druckmuller2008}. Monthly minimum averaged images are used as background images to remove the F-corona. They are accessible via the widely used SolarSoft IDL library. The resulting image can now be used to visualize features of interest such as bright coronal rays that are signatures of enhanced density structures. Streamer rays are good proxies to visualize the dense plasma emanating from the tip of helmet streamers, that in turn supplies the Heliospheric Plasma Sheet (HPS). Hence streamer rays are also useful to get an approximate location of the Heliospheric Current Sheet (HCS), which is likely located inside the thick HPS.

The second step is to concatenate and convert continuous observations into a synoptic format, that is heliographic latitude versus longitude at a given height in the corona. Such maps, if defined in the Carrington frame rotating with the Sun are called Carrington maps. They have been significantly used in the past to exploit 1 AU observations \citep[see e.g.][]{Wang1992,Wang1998,dePatoul2015,PintoRouillard2017,Sasso2019,Rouillard2020b}. 

To generate such a 2D map from LASCO C2 images, the following steps are followed. For 1 AU observations and small heliocentric distances, a widely used and reasonable approximation is to project images onto the helioprojective sphere (that is centered at the observer, also known as the plane-of-the-sky). The projected image is then converted from helioprojective to Carrington coordinates. As the Sun rotates on itself, the LASCO C2 coronagraph images the plane of the sky above both the East and West limbs, performing a full scan in half of the solar synodic period (~13 days). A 3D annular volume of white light intensity is therefore progressively filled up, providing a global overview of the structure of the solar corona at multiple altitudes, similarly to tomographic imagery techniques \citep{Morgan2020}. Finally, we extract a single slice from this 3D volume at a specific heliocentric distance, getting at last a 2D synoptic, or Carrington, white-light map. As long as new images are available, the Carrington synoptic map is continuously added to and updated, allowing the production of a near-real-time map. An example of a white-light synoptic map is shown in the middle panel of figure \ref{fig:metric2-illustration}, which has been extracted at 5.0Rs and combines SOHO LASCO-C2 images from October 30 to November 13, 2018.

Once the white-light synoptic map is complete, we perform 2-D median filtering in order to reduce ``salt and pepper'' noise while preserving edges of features of interest. The streamer belt is a thick band of enhanced brightness in each white-light synoptic map, that shows the overall shape of the HPS around the Sun. It is not always easily distinctive and sometimes can be confused with pseudo-streamers. One needs to adjust the height at which the white-light synoptic map is extracted in order to dispel the ambiguity. For instance, the usual larger radial extent of streamers compared to pseudo-streamers can help to discriminate between both \citep{Wang2007}. For a start, we selected a height of 5.0 Rs, large enough to remove most of the pseudo-streamer signatures, while also keeping good image quality \citep[further details can be found in ][]{Poirier2021}. The outermost boundary of the streamer belt is identified with a simple brightness threshold. The threshold has been adjusted manually until a coherent shape was obtained (see the dark contour in the middle panel of figure \ref{fig:metric2-illustration}). That provides a rough idea of the shape and thickness of the HPS. For this work, the key feature of interest is the HCS which is likely standing inside the streamer belt and can reasonably be expected to be approximately located where the brightness is maximum within the streamer belt. As such, we trace the locus of maximal brightness at each longitude (see the dark dashed line in the middle panel of figure \ref{fig:metric2-illustration}), hereafter called the Streamer-Maximum-Brightness (SMB) line via the method described in Algorithm box \ref{alg:metric2}. A good correlation has been found between the SMB line and the magnetic sector structure observed \textit{in situ} at 1 AU \citep[see][]{Poirier2021}. Similarities between the brightest regions of the K-corona and the magnetic sector boundary have also been highlighted in past studies \citep[][]{Hansen1974,Howard1974}.

\subsection{PSP and 1AU \textit{in situ} Magnetic Polarity Measurements}
\label{subsec:mag_pol}


The final observation we use in our model comparisons are direct \textit{in situ} measurements of the magnetic field polarity in the inner heliosphere over the time intervals of interest combined with knowledge of the corresponding spacecraft's position. We use data from the FIELDS instrument \citep{Bale2016} on the \textit{Parker Solar Probe} spacecraft \citep{Fox2016} down to 35.7$R_\odot$, as well as the OMNI ``near-earth'' dataset\footnote{\url{omniweb.gsfc.nasa.gov/}} and measurements from the IMPACT \citep{Luhmann2008} instrument on board \textit{STEREO AHEAD} \citep{Kaiser2008} at 1AU. The magnetic field polarity is determined by the sign of the radial component of the field ($B_R$). We consistently choose to determine this by computing the most probable value over 1 hour intervals of higher cadence data following \citet{Badman2020}. This removes small scale fluctuations and gives a resulting observable which indicates what side of the heliospheric current sheet (HCS) that specific spacecraft was located at a particular time. 
One further step is required for comparison with the location of the neutral line predicted by our coronal models. Since the coronal models predict the neutral line at a distance of a few solar radii, the spatial information of the spacecraft further out must be corrected to account for the Parker spiral. We utilize ballistic propagation \citep{Nolte1973,Macneil2021} whereby a constant solar wind velocity is assumed for a Parker spiral field line intersecting the spacecraft to set the spiral curvature. This velocity is chosen either from \textit{in situ} measurements of the velocity at the spacecraft or otherwise uses a compromise value of slow wind of 360 km/s (typical of low heliographic latitudes). By combining the polarity time series with the spatial location of the spacecraft, we can build up a picture of where warps in the HCS are located. An example of data from PSP/FIELDS during its first solar encounter (October/November 2018) is shown in figure \ref{fig:metric3-illustration} where, in the middle panel, we see PSP's trajectory (ballistically mapped to the edge of the corona) colorized by the polarity measured as it traveled through each position.

The final derived observation is then a list of timestamps, polarity values, and ballistically corrected heliographic latitudes and longitudes of the spacecraft at the outer boundary of the coronal model (typically 2.0-2.5 $R_\odot$ for PFSS models). Our method of comparison to this observation is described below in section \ref{subsec:metric3}.

\section{Prediction-Observable Comparisons}
\label{sec:pred-obs}

We next introduce our three metrics for comparing the data and modeled observables introduced in the previous sections. 

\subsection{Coronal Holes}
\label{subsec:metric1}
%

\begin{figure*}
   \centering
   \includegraphics[width=0.6\textwidth]{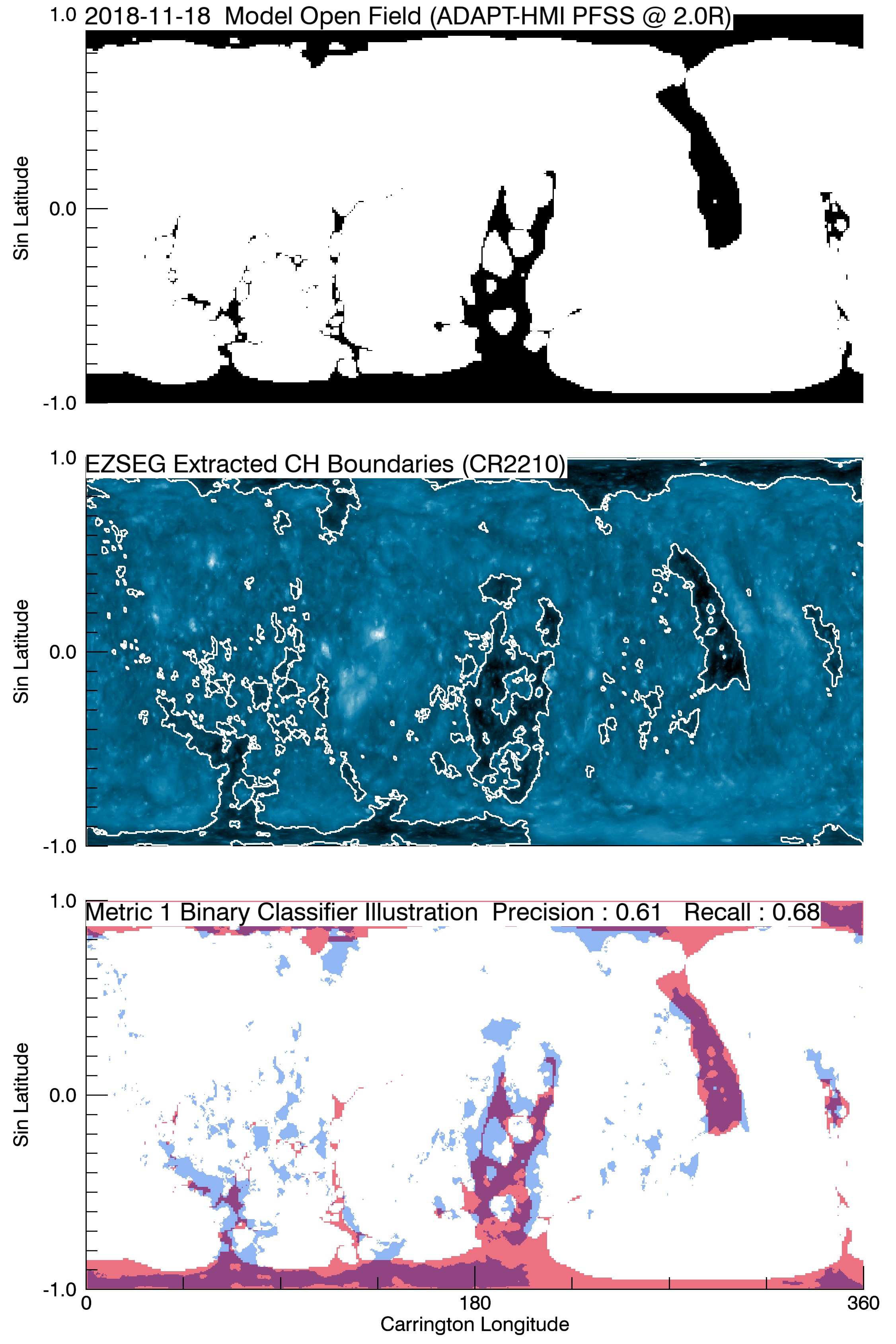}         
   \caption{Illustration of construction of metric 1 using the example of an ADAPT-HMI PFSS model from 2021/11/18 during PSP Encounter 1 with a source surface height parameter of 2.0$R_\odot$. Top panel: Mask of coronal hole locations (black shading) predicted by the PFSS mode model, i.e. pixels in the input magnetogram which map to open field lines. Middle panel: Coronal hole locations as determined by applying the EZSEG algorithm to AIA/EUVI maps of 193\AA~coronal emission, as described in the main text. The background image depicts the EUV Carrington map and the white contours indicate the areas identified by EZSEG. Bottom panel: Shading depicting the comparison of the observed and modeled coronal holes and metric construction. Blue shading indicates pixels that are identified as coronal hole by EZSEG, and red shading indicates locations predicted open by the model. Where they overlap, purple shading indicates the pixels are predicted and measured as coronal hole. The main text (section \ref{subsec:metric1}) describes how these regions are used to generate the precision, recall and final `f-score' metrics to quantify the agreement.}
   \label{fig:metric1-illustration}%
   \end{figure*}

\begin{table} 
\caption{Process to Generate Coronal Hole F-Score\label{alg:metric1}}
\centering
\begin{tabular}{| p{0.45\textwidth} |}
\hline
1) Prepare \citep[or download from Zenodo repository: ][]{Zenodo2022} a $193$\,\AA\, EUV Carrington map for time interval of interest.\\
2) Prepare coronal hole map from model output (0 = Closed, 1 = Open), Map A.\\
3) Apply EZSEG \citep{Caplan2016} to $193$\,\AA\, EUV Carrington map to obtain a binary map of ``measured'' open and closed pixels, again 0 = Closed, 1 = Open: Map B.\\
4) Interpolate model coronal hole map to the same resolution as the binary map\\
5) Compute true positive rate $t_p$ : Number of pixels open in Map A AND Map B\\
6) Compute false positive rate $f_p$ : Number pixels which are open in Map A and closed in Map B \\
7) Compute false negative rate $f_n$ : Number of pixels which are closed in Map A and open in Map B.\\
8) Follow equations \ref{eqn:binclass1}-\ref{eqn:binclass3} to compute precision, recall and f-score respectively. \\
\hline
\end{tabular}
\end{table}

Our aim here is to provide some measure of how well the coronal magnetic field models are able to predict the true open magnetic field on the Sun during each of the PSP encounters.  We compare the model predictions of the locations of open field (black regions, top panel, figure \ref{fig:metric1-illustration}) with the locations and extent of coronal holes detected in our 193\,\AA\, Carrington maps (white contours, middle panel, figure \ref{fig:metric1-illustration}). Each of the models outputs a Carrington map of open and closed magnetic field. Similarly, EZSEG (section \ref{ch_maps}) produces a pixel map of coronal hole detections. We take the EZSEG map of coronal hole detections to be the ground truth open field dataset and apply a binary classifier algorithm to make the model comparison. This overall procedure is summarized in algorithm box \ref{alg:metric1}.

The algorithm computes three metrics: the precision ($p$), recall ($r$), and f-measure ($f$). 
These are familiar concepts in machine learning and statistics and are defined as
\begin{eqnarray}
\label{eqn:binclass1}
p = t_p / (t_p+f_p) \\
\label{eqn:binclass2}
r = t_p / (t_p+f_n) \\
\label{eqn:binclass3}
f = 2 p r / (p+r)
\end{eqnarray} 
where $t_p$ is the number of true positives, $f_p$ is the number of false positives, and $f_n$ is the number of false negatives. Each metric falls in the range of 0--1.
Here, a true positive is recorded when a pixel marked as a location of open field in the model is found to be open in the ground truth map (purple or ``overlap'' pixels in the bottom panel of figure \ref{fig:metric1-illustration})
A false positive is recorded when a pixel marked as a location of open field in the model is found to be closed in the ground truth map (red pixels in figure \ref{fig:metric1-illustration}).
Similary, a false negative is recorded when a pixel location marked as closed field in the model is found to be open in the ground truth map (blue pixels in figure  \ref{fig:metric1-illustration}).
The precision is therefore the fraction of open field predicted by the model that is actually open according to the detection of coronal holes in the 193\,\AA\, Carrington synoptic map, that is the number of purple pixels as a fraction of the number of pixels which are red or purple. Conversely, recall is the fraction of actual open field (according to the coronal hole detection map) that
is predicted to be open by the model, or number of purple pixels as a fraction of the number of blue or purple pixels. These two metrics are useful for understanding how the models are performing, but, as can be seen in panel-by-panel comparison in appendix \ref{sec:appendixA}, they are anti-correlated, so we calculate the f-measure (the harmonic mean of the precision and recall) to classify which models are best. The harmonic mean gives high values if both constituents have high values, but gives low values if \textit{either} constituent is low-valued independent of the other.

By now it should be clear that there are some limitations to our comparisons. First, we assume that all the magnetic field within all the detected coronal holes is open. There is, of course, a transition to closed field close to the boundary. Second, we assume that there is no open field outside of coronal holes. Comparisons of PFSS extrapolations with actual solar features indicate that this is not the case - see, e.g., the full Sun intensity and open-field maps of \citet{Brooks2015} that suggest a large contribution to the slow solar wind from active region outflows. Although the time intervals studied in this work occur during deep solar minimum, we do have one interval when there was significant solar activity. 

As will be shown later, metric scores typically didn't exceed 60\%. Figure \ref{fig:metric1-illustration} illustrates some of the reasons for this. For the example shown, the model appears to overexpand the polar coronal holes and underexpand the mid-latitude ones. Since the polar coronal holes are the largest open regions by solid angle, they became an important point of comparison for the models and is what led us to prefer EZSEG as the algorithm for detection (recall the discussion
in Section \ref{ch_maps}). As we can see in the middle panel of figure \ref{fig:metric1-illustration} (white contours), EZSEG does a good job detecting the polar coronal holes while other
algorithms were less effective. 
Conversely, there are quite a few small observed coronal holes that appear dubious. These, however, 
constitute a small fraction of the total area of detections so are a minor contributor to our results. An example showing the f-measure comparisons for all the models we investigated for encounter 1 is included in the Appendix (Figure \ref{fig:AppendixA_Metric1_E1}) and linked to full encounter movies in the electronic version of the manuscript.

\citet{Wagner2021} also make comparisons between EUV images and open field models (specifically from the EUHFORIA model). In their work, they use a simple threshold condition after \citet{Krista2009} to obtain observed coronal holes, and use a similar set of binary classification metrics \citep[see][for details]{Wagner2021}. This work addresses the difficulty with polar coronal holes by truncating their binary classification to between $\pm60^o$ latitude, and therefore score the model based on mid and low latitude coronal holes only. In the present work, our choice is to use a more complex coronal hole extraction software (EZSEG) and therefore a slightly more convoluted analysis pipeline with the advantage of having a higher degree of confidence in our extracted polar coronal holes. 
Indeed, we should note that the polar coronal holes, in general, are poorly observed both in EUV and the magnetograms driving the models. However, their equatorward extent is usually well observed, at least in one hemisphere. Accepting this uncertainty, we make use of sine latitude (or cylindrical-equal-area, CEA) projections in our maps which limit the area occupied by the polar coronal holes, effectively assigning a lower weighting to those pixels by preserving pixel area at higher latitudes. Future work may seek to improve equations (\ref{eqn:binclass1}-\ref{eqn:binclass3}) with an explicit weighting for different types of coronal holes, and even further afield, direct solar polar observations may negate such considerations altogether. However, in the present work we proceed with this straightforward classification scheme and argue that due to the deep solar minimum conditions that our studied time intervals occur during, it is important to assess the polar coronal holes in our metrics.

\subsection{Streamer Belt}
\label{subsec:metric2}

\begin{figure*}
   \centering
   \includegraphics[width=0.6\textwidth]{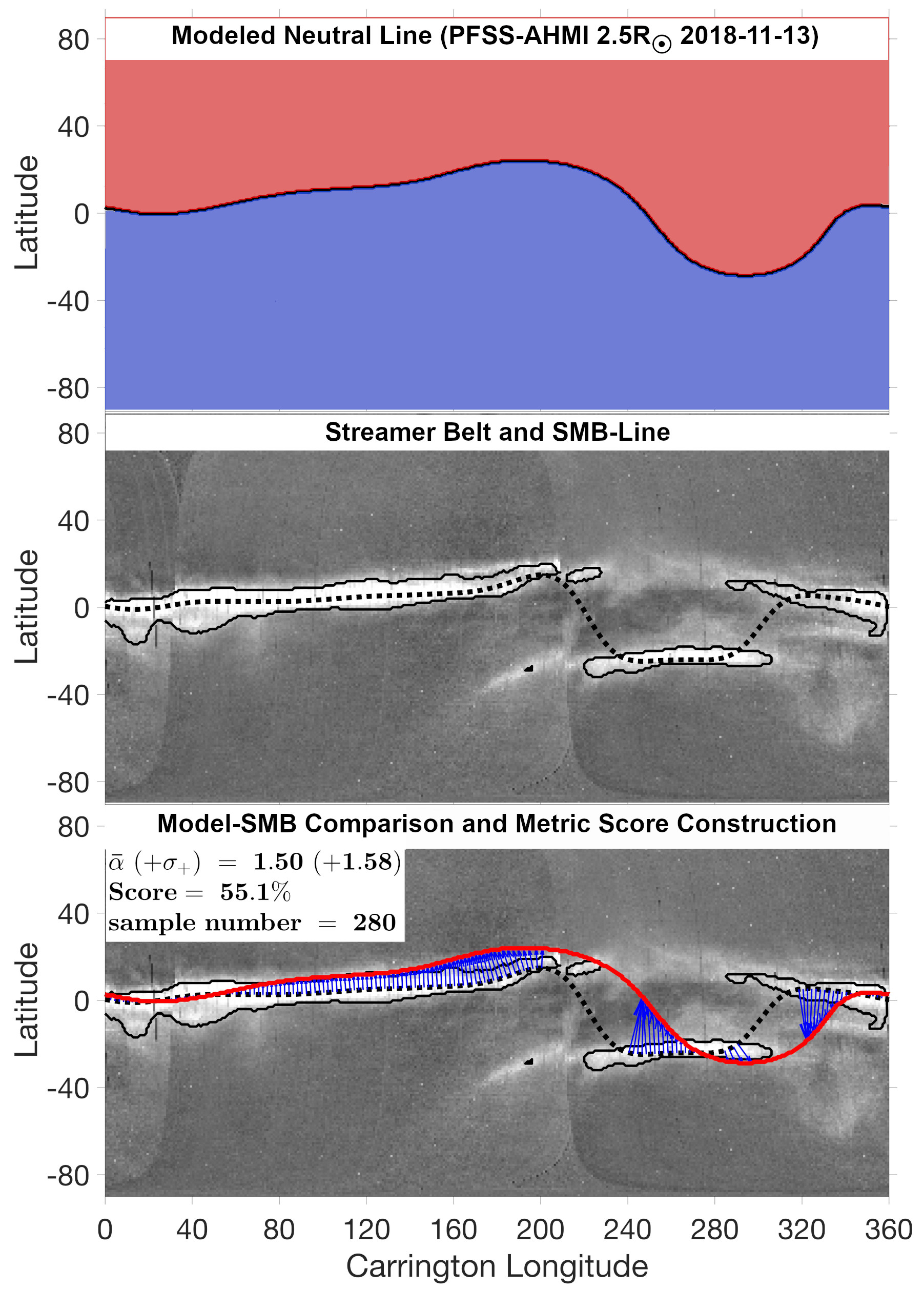}
   \caption{Illustration of the construction of Metric 2. Top panel: Model input - contour of polarity inversion line as predicted by coronal models. Middle panel : Observation input - a white light synoptic map of the streamer belt (Carrington map) constructed at a height of $5\ R_\sun$ from SOHO LASCO-C2 images taken between 2018, October 30 and 2018, November 13. The streamer belt contour is plotted in dark. The core of the streamer belt identified by the SMB line is traced as a dark dashed line. Bottom panel - Comparison between the SMB line and the modeled polarity inversion line (solid red line). The angular deviations between the two lines that are used to compute the white light (WL) metric are represented by the blue arrows. Results from the WL metric are given in the legend, of which the WL global score (in \%) calculated from the mean $\bar{\alpha}$ and mean deviation $\sigma_+$.}
   \label{fig:metric2-illustration}%
\end{figure*}

\begin{table}
\caption{Process to Generate White Light Streamer Belt Score \citep{Poirier2021}\label{alg:metric2}}
\centering
\begin{tabular}{|p{0.45\textwidth}|}
\hline
1) Prepare \citep[or download from Zenodo repository:][]{Zenodo2022} WL Carrington map for time interval of interest.\\
1a)(recommended) Clean up the WL map with a 2-D gaussian filtering (e.g. using the \verb+medfilt2()+ function in MATLAB). \\
2) Prepare neutral line from model output (2D map of polarity at model outer boundary, pixel values = $\pm1$). \\
3) Extract (or download) the SMB line. If downloaded then jump directly to step 4. \\
3a) Find longitudes and latitudes of max brightness points in the WL Carrington map. \\
3b)(optional) Smooth out the SMB line with a 1-D gaussian filtering (e.g. using the \verb+smoothdata()+ function in MATLAB). \\
4) Measure (or download) the streamer belt (half) width ($T_i$) (provided in the SMB .ascii files in the Zenodo repository). One could also simply assume a constant streamer belt width and jump to step 5. If you choose the download option or want to perform your own measurements, please read carefully the footnote \footnote{\label{WL_footnote}In this work and in the SMB files provided online \citep[Zenodo repository : ][]{Zenodo2022}, all angular distances are measured in the direction perpendicular to the SMB line, which requires to compute normal vectors beforehand. For simplicity one could choose to measure these angular distances over latitude (i.e. along the vertical axis) or just retain the closest distance to the streamer belt contour. Two streamer widths are given in the SMB files, for the northern and southern portion of the streamer belt, of which only one is used depending on the relative position of the modeled neutral line with the SMB line. One could also simplify the procedure by computing an average of the northern and southern (half) widths.}. Follow steps 4a and 4b if you decide to perform your own measurements or else jump to step 5. \\
4a) Draw streamer envelope with a simple contour threshold. \\
4b) Measure the angular distances ($T_i$) between each point of the SMB line and the extracted contour of the streamer envelope. \\
5) Measure for each grid point along the SMB line, the angular distance ($\alpha_i$) to the model neutral line extracted at step 2. Make sure to make these measurements along the same directions as in step 4b. \newline
6) Normalize each angular distance ($\alpha_i$) computed in step 5 by its associated streamer (half) width ($T_i$) measured in step 4. \\
7) Apply equation \ref{eq:WL_score} to obtain the streamer belt metric score $C(\bar{\alpha},\sigma_+)$. \\
\hline
\end{tabular}
\end{table}

As discussed in subsection \ref{subsec:WL_Maps_subsection}, we built and exploited white-light synoptic maps to get an approximate shape of the HCS, or magnetic sector structure, via what we called the streamer-maximum-brightness or SMB line. As a consequence the SMB line can then be easily compared to any global coronal model. The steps we follow are summarized in algorithm box \ref{alg:metric2}.
However with the SMB being extracted at a specific height, some precautions must be taken prior any comparison being made.

In this work, we performed PFSS modeling with a source surface height up to 3.0Rs. Above this height, the magnetic field is assumed to be purely radial and hence the line of polarity inversion or neutral line (NL) remains the same in a latitude vs longitude format. Assuming that the streamer topology does not change significantly between the PFSS source surface height and 5$R_\odot$, the SMB can be compared with the PFSS neutral line. The overall shape of the streamer belt is not expected to vary significantly but a slight narrowing of the streamer belt could be observed still. The outer boundary of the WSA runs is located at 5$R_\odot$, so at the exact same height at which the SMB line is extracted allowing a direct comparison. The MAS MHD simulations from PSI extend much further and the NL is extracted at 30$R_\odot$. In principle, this means the HCS could be shifted up to 8 degrees due to Parker spiral curvature, but in practise this is a small correction compared to the dispersion of neutral line topology amongst our dataset of model runs.

In order to compare in a statistical manner white-light observations and global coronal models we need to define a metric that is accurate and robust. To compare the SMB with the modeled neutral lines, various techniques can be tested. We decided to measure the angular distances between the two lines. These angular distances $\alpha_i$ are then normalized by the local streamer (half) thickness $T_i$ which is obtained from the streamer belt contour identified beforehand. The reason for choosing this normalization is twofold: the streamer thickness is a physical measure that helps to interpret the final metric; and also the final metric is therefore not affected by the streamer thickness variability over space and time. In particular, this normalization leads to the useful intuition that scores above 50\% correspond to model neutral lines lying on average within the bright white light emission of the streamer belt. The final step consists of calculating an averaged score over the full Carrington longitude range. The global score is built so that a model having a good overall match with WL coronagraphic observations would still be penalized in case of local and large deviations from the SMB line. Further details on the derivation of this global score can be found in \citet{Poirier2021} where the following formula has been retained:

\begin{equation}
    \begin{aligned}
    C(\bar{\alpha},\sigma_+) &= g(\bar{\alpha})\times [1-p(\sigma_+)] \\
    g(\bar{\alpha}) &= \frac{1}{1+\left(99\right)^{\bar{\alpha}/2-1}} \\
    p(\sigma_+) &= \frac{1}{1+\left( \frac{1}{99}\right)^{\sigma_+/2-1}} \\
    \end{aligned}
    \label{eq:WL_score}
\end{equation}

where $\bar{\alpha}=\frac{1}{n}\sum_{i=1}^n |\alpha_i|/T_i$ is the mean over longitude of the normalized angular distances to the SMB line, and $\sigma_+= \sqrt{\frac{1}{n}\sum_{i=1}^{n}(|\alpha_i|/T_i-\bar{\alpha})^2}$ represents the mean deviation. Note that a looser version of the formula has been adopted in this work to better cover the dispersion amongst all model runs.

  
An example is illustrated in Figure \ref{fig:metric2-illustration} where the WL metric has been applied to a PFSS model based on the ADAPT/HMI map of 2018, November 13. For this particular case, the polarity inversion line remains mostly at the edge of the streamer belt that is reflected with a mid-range WL score of $55\%$. In regions where the HPS is significantly warped, the WL signature is too faint to allow a precise detection of the shape of the streamer. This happens when the line-of-sight of an imager, or coronagraph, gets significantly inclined with respect to the HPS such that the light scattered by electrons is integrated over a much shorter path. Hence, the WL metrics is configured to not evaluate the model in these specific regions, as shown by the absence of blue arrows in Figure \ref{fig:metric2-illustration}. This constrains the effectiveness of this metrics to simple solar configurations where the HCS shape is not too complex. As a consequence, the reliability of this metric is uncertain for highly structured coronal configurations typically observed during periods of high solar activity. Future improvements which may increase the reliability of this metric for more complex configurations are discussed in further detail in \citet{Poirier2021}. The conversion in an all-in-one metric that evaluates both the pseudo-streamers and regular streamer locations should alleviate the uncertainty in using the SMB line alone for complex coronal configurations. WL synoptic maps produced by merging multi-viewpoint coronagraphic observations \citep[see e.g.][]{Sasso2019} should also be more appropriate to catch the dynamic reconfigurations of the solar corona arising with higher solar activity.

\citet{Wagner2021} also define metrics to compare white light observations to model results, focusing on snapshots of coronagraph observations and projecting their model results into the relevant plane of the sky. Their method requires a user to ``point and click'' to associate features in model and data and measure, for example, the angular difference between a modeled and observed streamer structure. This method enables very detailed comparisons at different coronal altitudes, albeit at the cost of requiring human intervention, and also only focuses on one plane cut through the model at a given time. Our chosen method after \citet{Poirier2021} is automated and evaluates agreement at all longitudes, but at the cost of not tracking variation with coronal altitude.

\subsection{Magnetic Sector Structure}
\label{subsec:metric3}


\begin{figure*}
   \centering
   \includegraphics[width=0.7\textwidth]{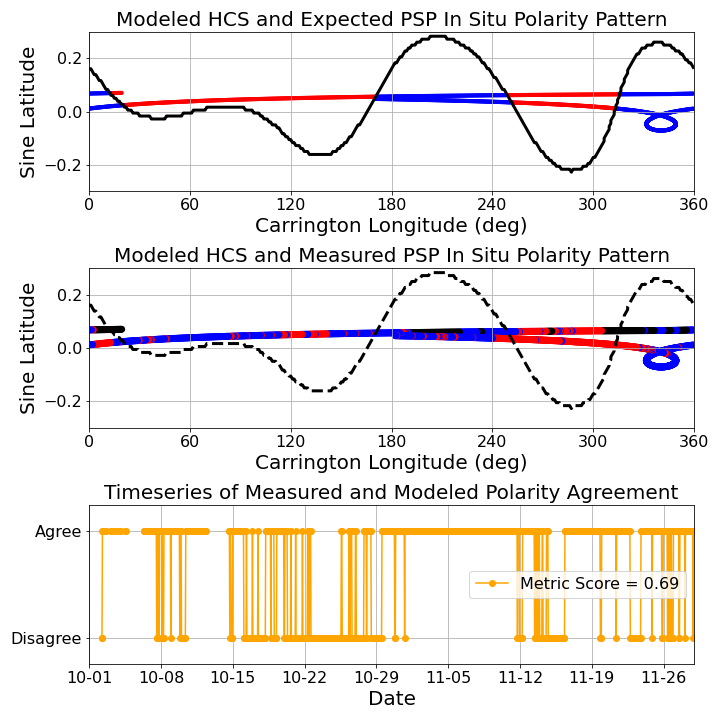}         
   \caption{Illustration of Construction of Metric 3. Top panel: Model input - contour of polarity inversion line as predicted by coronal models and resulting predicted polarity that PSP's trajectory is expected to trace out according to the model. Middle panel : Observation input - same trajectory of PSP but colored by measured polarity. Dashed curve indicates the same polarity inversion line from the top panel showing the model-data agreement is imperfect. Bottom panel: Product of measured and predicted  polarity timeseries. Values are +1 for agreement and -1 for disagreement. Metric is the ratio of agreements to the number of measurements, as described in the main text, Section  \ref{subsec:metric3} }
   \label{fig:metric3-illustration}
\end{figure*}

\begin{table}
\caption{Process to Generate In Situ Polarity Score \label{alg:metric3}}
\centering
\begin{tabular}{|p{0.45\textwidth}|}
\hline
1) Prepare \citep[or download from Zenodo repository: ][]{Zenodo2022} measured in situ polarity timeseries. If preparing yourself, recommended to follow processing routine here where for each hour of data, generate histogram of $B_R$ and take polarity of most probable value as the polarity ``measurement'' for that time. \newline
2) Produce model-predicted polarity timeseries by following these steps : \\
2a) Use spice kernels to generate spacecraft position in the heliographic frame at an hourly cadence over the time interval of interest. \\
2b) Ballistically propagate this position inwards to 2.5$R_\odot$ using the equation for the Parker spiral \citep[Equation 1, ][]{Badman2020} \\
2c) Sample the modeled 2D map of the neutral line using this projected orbit (via interpolation or nearest neighbor) to generate a timeseries of predicted polarity.\\
3) Combine modeled and measured timeseries via equation \ref{eqn:metric3} to obtain the in situ polarity metric score. \\
\hline
\end{tabular}
\end{table}

From our \textit{in situ} observations described in Section \ref{subsec:mag_pol} we have a measured timestamp, polarity and a ballistically mapped spacecraft location at the outer boundary of a given coronal model. To compare this observation to the model's prediction of the location of the heliospheric current sheet, we produce a predicted time series of magnetic polarity. For each timestamp, we take the model output closest in time to that timestamp and sample the polarity at the outer boundary of the model at the location of the associated ballistically mapped footpoint location. We then have for each timestamp, $i$, a measured ($m$) and predicted ($p$) value of the polarity. While each individual measurement-prediction comparison gives very little information, we can multiply the whole timeseries together, and divide the total number of agreements (value of +1) by the total number of measurements to give a single value representing the fraction of measurements and predictions which agree, giving a single number between 0 and 1 evaluating the model and \textit{in situ} measurement agreement for a given encounter. This operation may be mathematically expressed as with the Heaviside step function $\Theta(x)$ as :

\begin{equation}
    \centering
    \sigma(\text{sign}(B_R)) = \frac{1}{N}\sum_{i=1}^N\Theta(B_R^{m,i}B_R^{p,i})
    \label{eqn:metric3}
\end{equation}

We note that a value of this metric of 0.5 would indicate as many agreements as disagreements in the timeseries and thus approximately equivalent to random chance. Thus, values above $0.5$ are required for this metric to convey ``good'' agreement. 

The construction of this metric is illustrated in Figure \ref{fig:metric3-illustration} in a similar 3 panel format as for the previous two metrics. In the top panel, a modeled HCS is depicted as a solid black curve and the trajectory of PSP projected onto the model outerboundary is superimposed. The color of the trajectory (red or blue) shows the model's prediction of the \textit{in situ} polarity measurement according to which side of the HCS PSP is located. The equivalent data is shown in the middle panel where the same trajectory is colorized by the measurements for the relevant time interval. For reference, the model HCS is again included as a dashed curve to show that the measurement and model do not agree perfectly. Finally, the construction of the metric is illustrated in the bottom panel where a timeseries of predicted and observed \textit{in situ} measurements are combined to produce a timeseries of agreement or disagreement. The score shown in the metric is produced from this data according to equation \ref{eqn:metric3}. This overall procedure is also summarized in algorithm box \ref{alg:metric3}.

\section{Results}\label{sec:results}


Having introduced our models, derived observables, sources of observation and the construction of each metric in the prior sections, we proceed to compute metric scores and explore the results. We study time intervals for +/- 30 days for each of PSP's first three encounters, and for each encounter study metric scores for PFSS models with different source surface heights ranging from 1.5$R_\odot$ to 3.0$R_\odot$, and with different magnetogram inputs (ADAPT-GONG, ADAPT-HMI and GONGz), and for WSA and MAS models for the same time intervals. Given the large number of parameters to explore (time interval, magnetogram, source surface height, model physical content), we divide the presentation of the results into a series of specific studies. A more exhaustive exposition of the scores as plots and movies are shown in appendix \ref{sec:appendixA} and online material.

\subsection{PFSS Models : Source Surface Height Study}
\label{subsec:results1-vs_rss}

\begin{figure*}
   \centering
   \includegraphics[width=0.8\textwidth]{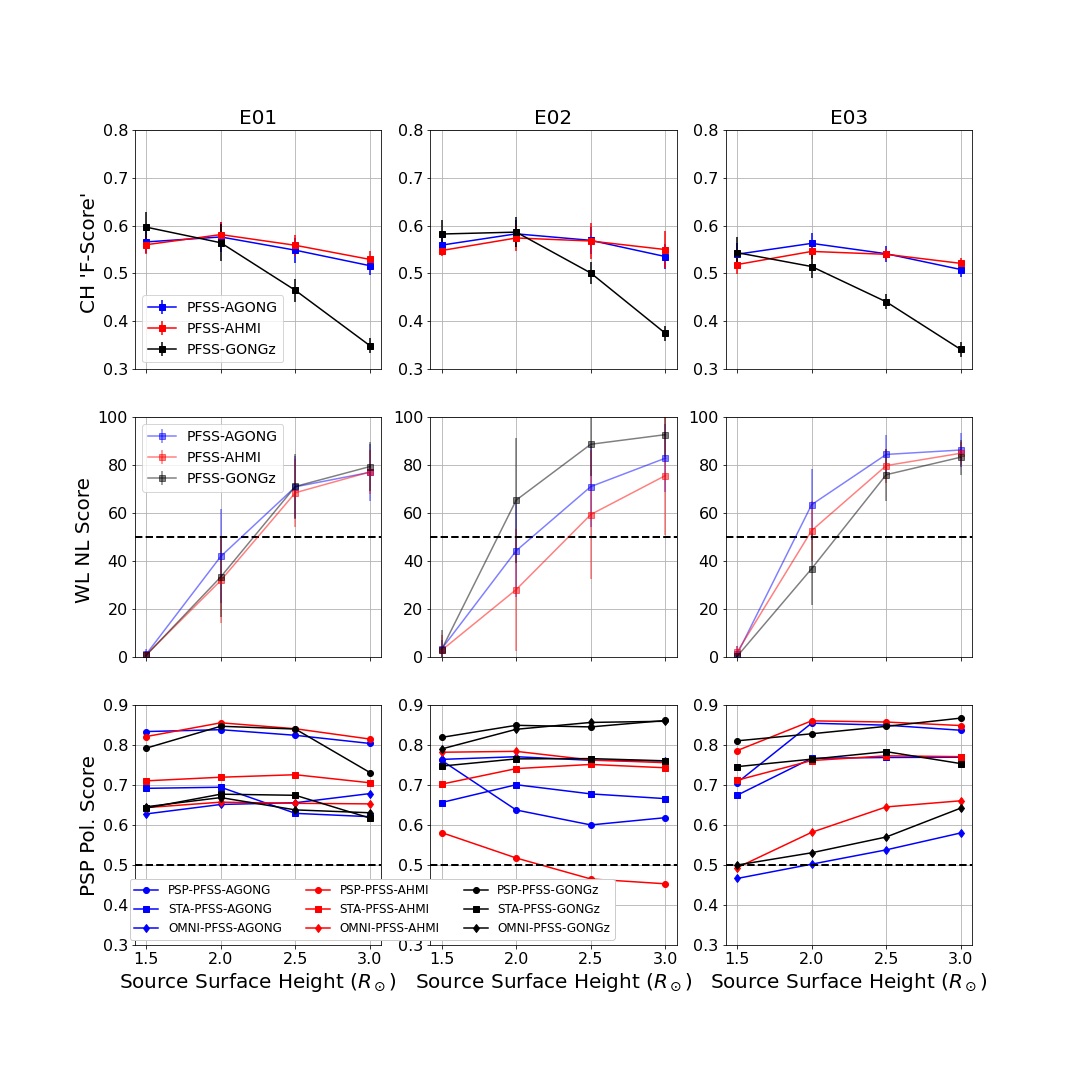}         
   \caption{Summary of metric scores for PFSS models as a function of source surface height. Left, middle and right column show PSP Encounters 1,2 and 3 respectively. Top, middle and bottom row show metric 1 (coronal holes - Section \ref{subsec:metric1}), metric 2 (streamer belt topology - Section \ref{subsec:metric2}) and metric 3 (\textit{in situ} magnetic polarity - - Section \ref{subsec:metric3}). Each panel shows the respective metric score (averaged over the corresponding PSP encounter for metrics 1 and 2) as a function of source surface height of the PFSS model. Different colors indicate results utilizing different magnetograms (Blue : ADAPT-GONG, Red : ADAPT-HMI, Black: GONGz). The bottom row showing metric 3 has \textit{in situ} polarity scores for PSP/FIELDS, STEREO A/MAG and OMNI data shown by different symbols (circles, squares and diamonds respectively). All metrics are oriented such that a higher metric value indicates a ``better'' score. The middle row depicting metric 2 has a dashed horizontal line marking the 50\% score threshold which indicates that the modeled HCS lies primarily within the observed extent of the streamer belt. The bottom row has a similar horizontal dashed line indicating the 0.5 score threshold corresponding to the result randomly generated measured polarities would achieve. Metric 1 and 2 scores have error bars indicating the standard deviation of the score throughout the encounter.}
   \label{fig:results1-vs_rss}
\end{figure*}

We begin by studying metric scores for the Potential Field Source Surface models. In figure \ref{fig:results1-vs_rss}, metric scores are depicted for each encounter (column by column) as a function of source surface height. Results using different input magnetograms are shown in different colors (blue for ADAPT-GONG, red for ADAPT-HMI and black for GONGz). For metrics 1 and 2 the score shown is the average over that encounter, and the error bar is the standard deviation. For metric 3, there is only one value as the metric is integrated over the encounter and there is therefore no equivalent error bar. 

The coronal hole metric is shown in the top row of the figure. All models depict a lower score for the higher source surface heights and have the same qualitative variation for each encounter. The two ADAPT magnetograms (blue and red) follow each other very closely, with the GONGz diverging in qualitative behavior: The ADAPT magnetograms consistently show a maximum coronal hole F-score for a source surface height of 2.0 $R_\odot$ with a small reduction at either lower or higher values. GONGz magnetograms result in a maximum F-score for the lowest source surface height (1.5$R_\odot$) and a monotonic decrease for higher values and gives significantly lower values for 2.5 and 3.0$R_\odot$ source surface heights. The actual average scores vary with a maximum of 0.6 and a minimum for GONGz of $\sim$0.3 and for the ADAPT maps around 0.55. Error bars indicate the variation of the metric over time within each encounter is small ($<$10\%), which will be verified later in section \ref{subsec:results34-vs_time}. The bifurcation between the coronal hole representation of the ADAPT maps compared to GONGz, independent of which time interval is examined, suggests the surface flux transport treatment of the ADAPT maps has a stronger impact than variation between GONG or HMI data. Specifically, the ADAPT scores optimize around 2.0$R_\odot$ while for GONGz, the lower the source surface height the better. As discussed further in section \ref{sec:discussion}, this appears to relate to the extent of the polar coronal holes in the model and therefore may relate to the treatment of polar flux. The worsening scores with increasing source surface height is driven by the smaller extent of coronal holes predicted by PFSS as the outer boundary of the model is raised \citep[e.g.][]{Lee2011}. The maximum F-score derived is around 60\% which is far from perfect. As a reminder, the F-score is the harmonic mean of the precision and recall, which have the physical meanings respectively of "fraction of measured open field captured by the model" and "fraction of modeled open field which is actually open". 

Results for the streamer belt metric are depicted in the middle row of figure \ref{fig:results1-vs_rss}. Conversely to the coronal hole metric, the PFSS representation of the streamer belt according to our metric \textit{improves} monotonically with source surface height. This is true for each encounter, and independent of the magnetogram driving the model. The scores vary from near zero for the lowest source surface heights (1.5$R_\odot$) to greater than 80$\%$ for the highest values. The improvement with source surface height starts to flatten above 2.5$R_\odot$. A dashed curve at a metric score of 50\% indicates the score at which the majority of the modeled HCS remains within streamer belt latitudes, as defined by the intensity threshold discussed in Section \ref{subsec:metric2}. All PFSS models exceed this score for $R_{SS} \ge 2.5R_\odot$ and many exceed it by $2.0R_\odot$. There is variation in the ordering of scores according to magnetogram from encounter to encounter: For encounter 1 all model scores are nearly on top of each other with ADAPT-GONG being slightly better at $R_{ss}=2.0R_\odot$. For encounter 2, there is significant divergence, with GONGz doing the best followed by ADAPT-GONG and ADAPT-HMI. For encounter 3, GONGz has the lowest score, with ADAPT-HMI slightly better and ADAPT-GONG the best. The monotonic increase in score with source surface height is driven by the extent to which the modeled HCS is warped: At lower values of $R_{SS}$ PFSS derived HCS curves are very warped and span large ranges in latitude, while at higher values the HCS flattens out (tending to a flat line at infinite $R_{ss}$). Clearly, the observed latitudinal extent of the streamer belt is smaller than the HCS variation produced by PFSS models for lower source source surface heights.

Finally, the bottom row shows the \textit{in situ} polarity scores. In addition to the difference between different models, results are shown for \textit{in situ} data taken by PSP from $\sim$ 37.5-50$R_\odot$ (circles), and at 1AU by OMNI (diamonds) and STEREO A (squares). A dashed curve at a score of 0.5 indicates the threshold above which the metric score is better than random chance; the majority of scores do exceed this value. In general, the scores are relatively flat with respect to source surface height. Interestingly, although they are nominally testing the same model output as the streamer belt metric (model HCS), the profiles are more similar to the coronal hole metric. The highest metric scores are typically largest for 2.0 or 2.5$R_\odot$ with a small drop off for the low and high source surface height extremes. In terms of comparison between spacecraft, for encounters 1 and 3, PSP gave the highest scores for all source surface heights independent of magnetogram. For encounter 2 PSP gave both the lowest scores (ADAPT-HMI near 0.5) and the best (GONGz near 0.75). Other than for encounter 1, it is not clear that PSP's \textit{in situ} measurements close to the Sun are better modeled than \textit{in situ} measurements at 1AU. For encounter 3, PSP and STA performed identically (with high scores) while OMNI data gave significantly worse scores near the random chance threshold. The variation between spacecraft metric scores and different magnetogram scores for the same spacecraft is not consistent. For encounters 1 and 3, scores for a given spacecraft are tightly clustered together for different magnetograms while for encounter 2 there can be wide variation in a given spacecraft score for different magnetograms.

Taken together, there is a clear tension between our different metrics. The higher the source surface height, the better agreement is produced between the modeled HCS and the observed streamer belt location, while at the same time the coronal hole extent produced by the model decreases and gives worsening coronal hole scoring. The higher source surface height also gives slightly worsening \textit{in situ} scores, implying that while the amplitude of warping of the HCS is more consistent with streamer belt observations, the locations where those warps intersect spacecraft orbits are less well represented. These results demonstrate that there is no global optimum source surface height which gives a best representation of all aspects of the magnetic structure of the corona. 

To investigate if this tension can be eased by the physical insight introduced in more complex coronal models, we compare PFSS results to those of other models in the next section.

\subsection{Inter-model Comparison}
\label{subsec:results2-vs_model}

\begin{figure*}
   \centering
   \includegraphics[width=0.7\textwidth]{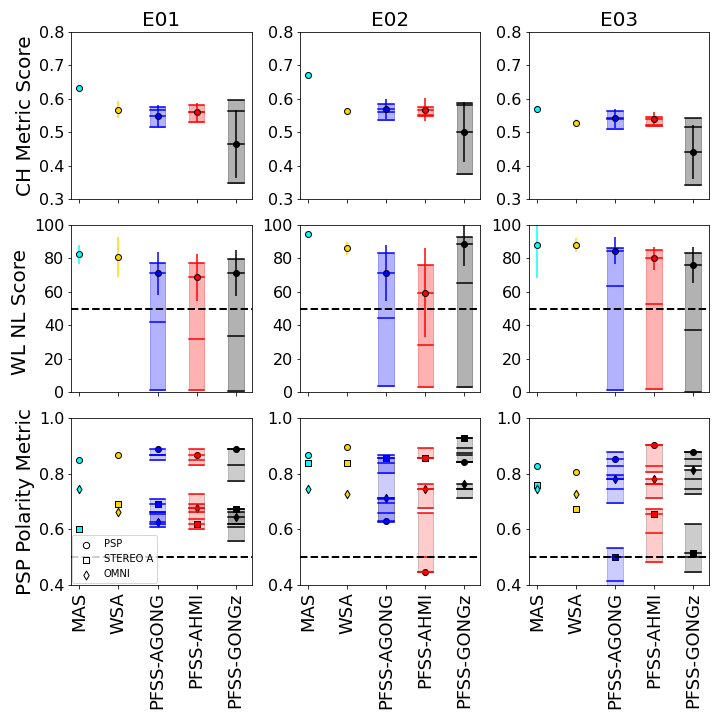}
   \caption{Inter-model comparison of metric scores. Columns and rows arranged as in figure \ref{fig:results1-vs_rss} separating out results for each metric and PSP encounter. In each panel, results are shown for the MAS model (cyan), WSA model (yellow) and PFSS models driven by ADAPT-GONG (blue), ADAPT-HMI (red) and GONGz (black) magnetograms. Mean and standard deviation in metric scores across each encounter are indicated by the plot markers and error bars. For the PFSS models, the plot marker and error bar indicates results for a source surface height of 2.5$R_\odot$; shading and horizontal bars indicate the range of metric scores with variation in source surface height. For metric 3 (bottom row), circle, square and diamond markers show different results for PSP, STEREO A and OMNI \textit{in situ} measurements respectively. Dashed horizontal lines in rows 2 and 3 are as in figure \ref{fig:results1-vs_rss}.}
   \label{fig:results2-vs_model}
\end{figure*}

Metric scores for the MAS MHD model, WSA model,  and the PFSS results discussed in the previous section are compared in figure \ref{fig:results2-vs_model}. The panels are laid out as in figure \ref{fig:results1-vs_rss} (columns and rows differentiating encounter number and metric respectively). In each panel, a metric score for each model is shown, with the main new information coming from the scores for the MAS model in cyan and WSA models in yellow. For the PFSS results, the main marker and error bar shows the results for a source surface height for the standard value of $2.5R_\odot$, but the variation within source surface height is captured by shaded regions indicating the range of variation, and horizontal bars indicating the distribution of scores which generate that range. The blue, red and black colors follow the scheme from figure \ref{fig:results1-vs_rss} in differentiating ADAPT-GONG, ADAPT-HMI and GONGz magnetograms respectively. 

The top row again shows the coronal hole metric scores. For each encounter the MAS model produces higher scores than the best achievable score for PFSS (although only slightly better for encounter 3). There is no error bar shown for the MAS score since there is only 1 coronal hole map generated due to computational tractability. As a reminder, the MAS model is driven by magnetograms from the Carrington rotation of the perihelion in each encounter (CR2210,CR2215 and CR2221).  WSA on the other hand does not show significant improvement over the PFSS models in terms of coronal hole representation. For encounter 1, it is slightly better than the 2.5$R_\odot$ ADAPT PFSS results but less than the best score generated by lowering the source surface height. For the other encounters it is slightly lower scoring than 2.5$R_\odot$ ADAPT PFSS results. Note that the WSA model is driven by ADAPT-GONG magnetograms.  

Metric 2 as depicted in the middle row of figure \ref{fig:results2-vs_model} shows that typically the MAS and WSA models produce better agreement between the model HCS and the observed extent of the streamer belt than the best achieved by PFSS results (recall these best scores were obtained by the highest source surface height explored, 3.0$R_\odot$), with the exception of the encounter 2 PFSS-GONGz results which produced comparable scores to the other models for 2.5-3.0$R_\odot$ source surface height. The PFSS shaded bars in this row demonstrate the stark decline in PFSS scores with lowering source surface height, and we again see heights of 2.0$R_\odot$ or below tend to produce scores less than 0.5 (dashed horizontal line), indicating the model HCS is reaching higher latitudes than expected from streamer belt observations. In all cases, the MAS score is the same or higher than the WSA score. 

Metric 3 finally is compared between models in the bottom row of figure \ref{fig:results2-vs_model}. As previously, circles, squares and diamonds indicate scores for PSP, STEREO A and OMNI \textit{in situ} measurements respectively. For encounters 1 and 3, for each model the best score is for PSP and scores worsen for OMNI and STEREO A. The ordering of the spacecraft scores is consistent between metrics (PSP $\>$ STEREO A $\>$ OMNI for E01 and PSP $\>$ OMNI $\gg$ STEREO A for E03). Comparing scores for a single spacecraft between models, for E01 and E03, MAS and WSA models give comparable or slightly worse scores compared to PFSS. The exception is the STEREO A scores for E03 which are significantly better in the WSA and MAS models as compared to the PFSS scores which are close to the 0.5 random chance threshold.  

Staying with the \textit{in situ} polarity metric, the result for encounter 2 is significantly different: the ordering of the spacecraft scores is reversed between the MAS and WSA models compared to PFSS models: PSP continues to give the best score for MAS and WSA, but is the worst for the ADAPT PFSS models (middle scoring for GONGz). We also note the variation in the PFSS scores as a function of source surface is often comparable to the difference between different spacecraft at the same source surface height, indicating that the location in the heliosphere at which the \textit{in situ} data is taken can be as important as the specifics of the model. We note that since the boundary conditions for the WSA model in encounter 2 is a single magnetogram chosen specifically to match PSP polarity measurements, it is not surprising that this specific value of metric 3 is one of the highest observed.

Lastly, bringing the metrics together we make some observations about the differences between the models overall. We see that:

\begin{enumerate}
    \item Despite only utilizing a single magnetogram compared to the timeseries of magnetograms for the PFSS Models, the MAS model's use of MHD physics leads to improvements of both the coronal hole metric and the streamer belt representation. We note that quiescent solar minimum conditions may be responsible for the applicability of the MHD result from a single magnetogram over a long time interval. It does not significantly improve (or may even worsen) predictions of HCS crossing times (which drives the score of metric 3), except for the second PSP encounter specifically for PSP \textit{in situ} measurements.
    \item The WSA model improves the score of metric 2 (streamer belt shape) compared to raw PFSS models, but it does not lead to a noticeable improvement in the coronal hole metric score. It also does not significantly improve scores for the \textit{in situ} polarity metric, except for encounter 2 where this was specifically tuned for. As discussed in Section \ref{sec:discussion}, this is an unsurprising result based on the WSA physics which only adds information above the source surface, and therefore forms a good sanity check of our metrics.
\end{enumerate}

Next, we close our exposition of the metric score results by looking at the details of the metric score timeseries for each encounter, from which the averages and error bars in the previous plots were generated.

\subsection{Metric Score Timeseries}
\label{subsec:results34-vs_time}

As a reminder, for each encounter for metric 1 and 2, the model output (coronal hole map and neutral line/HCS map) for PFSS and WSA models were evolved daily by using a new magnetogram (excepting E02 for WSA). The observed White Light Carrington map was also updated daily as the longitudes of the streamer belt visible from Earth updated. This means that, for metrics 1 and 2, we have a timeseries of metric scores for the 60 days comprising each encounter for WSA and PFSS models. For metric 3, the score is the output of integrating the full 60 day ``measured'' and ``modeled'' polarity timeseries together and therefore is not itself a timeseries.

\subsubsection{Coronal hole and streamer belt metric timeseries}

\begin{figure*}
   \centering
   \includegraphics[width=0.8\textwidth]{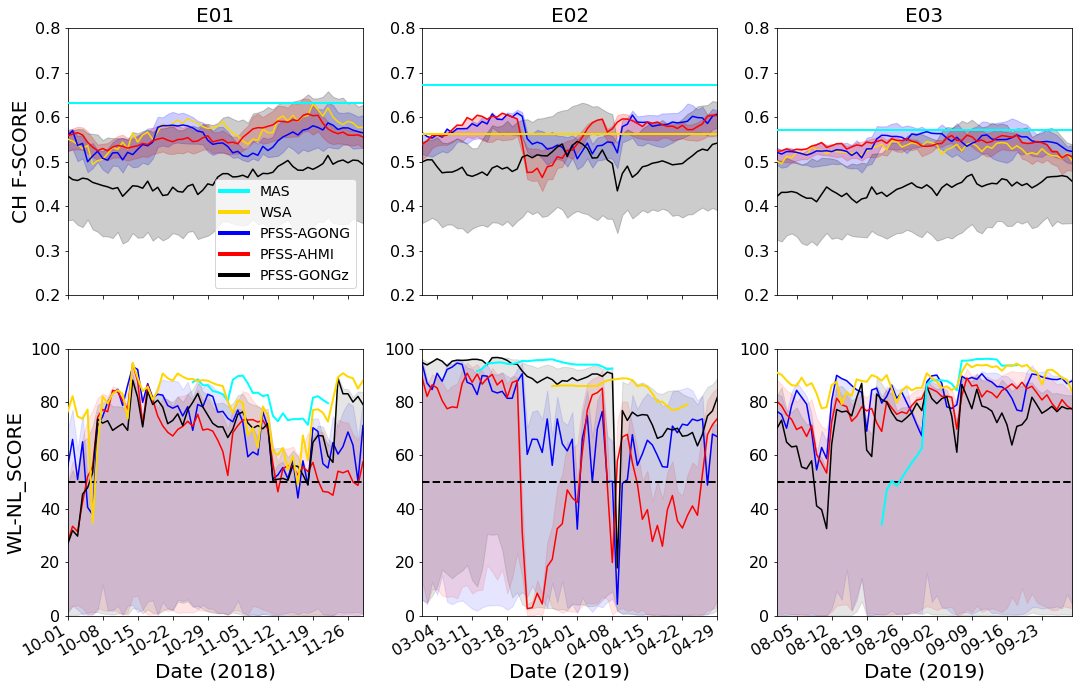}         
   \caption{Variation of Metric 1 and 2 scores as a function of time within each encounter at a daily cadence. In each panel, colored solid lines indicate different models with the same color scheme as in figure \ref{fig:results2-vs_model} where a source surface height of 2.5 $R_\odot$ is chosen for the PFSS models (blue, red, black). Variation according to source surface height for the PFSS models is indicated by the shaded regions bounded by the max and min metric score at each timestamp. The top panel shows the coronal hole score (metric 1), the bottom row shows the streamer belt score (metric 2). Note the timeseries for the MAS model (cyan) is a single value as there is only 1 magnetogram used for each encounter owing to the intractability of running 3D MHD models with time-dependent boundary conditions. For encounter 2, WSA also only uses a single magnetogram and so only has a single value for metric 1 (see main text, Section \ref{subsec:metric3}).}
   \label{fig:results3-12_vs_time}
\end{figure*}

In figure \ref{fig:results3-12_vs_time}, we depict the timeseries for metrics 1 and 2. The color scheme follows the preceding figures. The top row shows how the agreement between the modeled coronal holes and the EZSEG output of the EUV Carrington map for the relevant rotation evolved each day through each encounter. The bottom row shows the same for the agreement between the modeled neutral line and the observed streamer belt Carrington map. For the PFSS models, semi-transparent shading in the relevant color shows the range of metric values explored by varying the source surface height for each timestamp, while the solid lines show the scores for the canonical source surface height of 2.5$R_\odot$. The upper bound of these shaded regions therefore shows the timeseries of the optimized source surface height for each mode. Following these timeseries, we can see how the coronal hole representation and streamer belt representation of each model evolved with each encounter.

Examining the coronal hole representation (top row), we see for encounter 1 (top-left panel) of the ADAPT-driven PFSS models and the WSA model (also driven by ADAPT magnetograms) follow each other closely, showing a periodic variation peaking twice, with an amplitude of approximately 0.1 corresponding to about 10\% variation in the score. We note that the ``observation" EUV Carrington map is static in each time interval, utilising data combined over the course of a Carrington rotation, therefore all evolution shown derives from the evolution of the PFSS model driven by updating its magnetogram boundary condition each day. The PFSS-GONGz score (the only daily-updated model driven by a non-ADAPT magnetogram) is quite different for the same source surface height (2.5$R_\odot$), being much lower scoring the entire time, and not following the same periodic pattern. Its variation as a function of source surface height (grey shaded region) does show that at low source surface height, the GONGz maps can still produce good coronal hole scores. Some intervals in fact show the GONGz optimized score to be the best performing of the PFSS models, most notably during the ``active region emergence'' phase of encounter 2 (see Section \ref{sec:discussion} conclusion C2).

This bifurcation of scores can be seen in the other encounters coronal hole metric scores. For encounter 2, the timeseries are approximately separated into thirds, with the first and last third corresponding to similar scores as encounter 1 for the ADAPT driven models, while in the middle third the scores suddenly drop to similar scores as compared with the GONGz models. Encounter 3 maintains the bifurcation the whole time and all profiles are very flat. GONGz scores are lower the whole time while the other maps scores are indistinguishable. Additionally, the MHD score (which is not time varying due to the single model run and single EUV observation) may be compared to the time varying scores via the horizontal cyan line. For all encounters, the score is never exceeded by the best PFSS (2.5$R)\odot$) or WSA score. For encounter 2 it is better than all other models and input parameters. In encounters 1 and 3 there are time intervals where different (lower) source surface heights yield scores comparable to or exceeding the MHD value, as indicated by the upper extent of the shading in the figure. 

Next, examining the white light metric (bottom row), there are some qualitative similarities and differences to the behavior of the coronal hole scores. For encounters 1 and 3, all models (now including GONGz) follow similar variation in time. The encounter 1 time variation does not show the same periodicity as for the coronal hole metric, instead showing a flat, higher score for the first two-thirds of encounter 1, followed by a sharp drop in score (less pronounced for the MAS model). For encounter 3, a similar drop is observed near the start of the time interval followed by a higher and flat score for the remainder. However, these scores should be interpreted carefully during the first third of encounter 3 since significant portions of the streamer belt were missing due to an issue with the background subtraction (see the animated version of figure \ref{fig:AppendixA_Metric2_E1} for E3 in the online material). Regions with such artifacts are excluded from the metric computation but deteriorates the reliability by decreasing the number of measurements. Encounter 2, is again distinct from 1 and 3. As with the coronal hole metric, there are distinct thirds of the encounter with the first third having all models having high scores indicating agreement with the observations, followed by a precipitous drop in some models, and a further precipitous, but brief (1-2 days only) drop in all models. It is clear there are sharp changes detected by both metrics in encounter 2 on approximately March 23rd 2019 and April 9th 2019. As will be explored in Section \ref{sec:discussion}, encounter 2 did contrast to encounters 1 and 3 in terms of the presence of dynamic solar activity through that time interval.  

\subsubsection{\textit{In situ} Polarity Timeseries}

\begin{figure*}
   \centering
   \includegraphics[width=0.8\textwidth,angle=0]{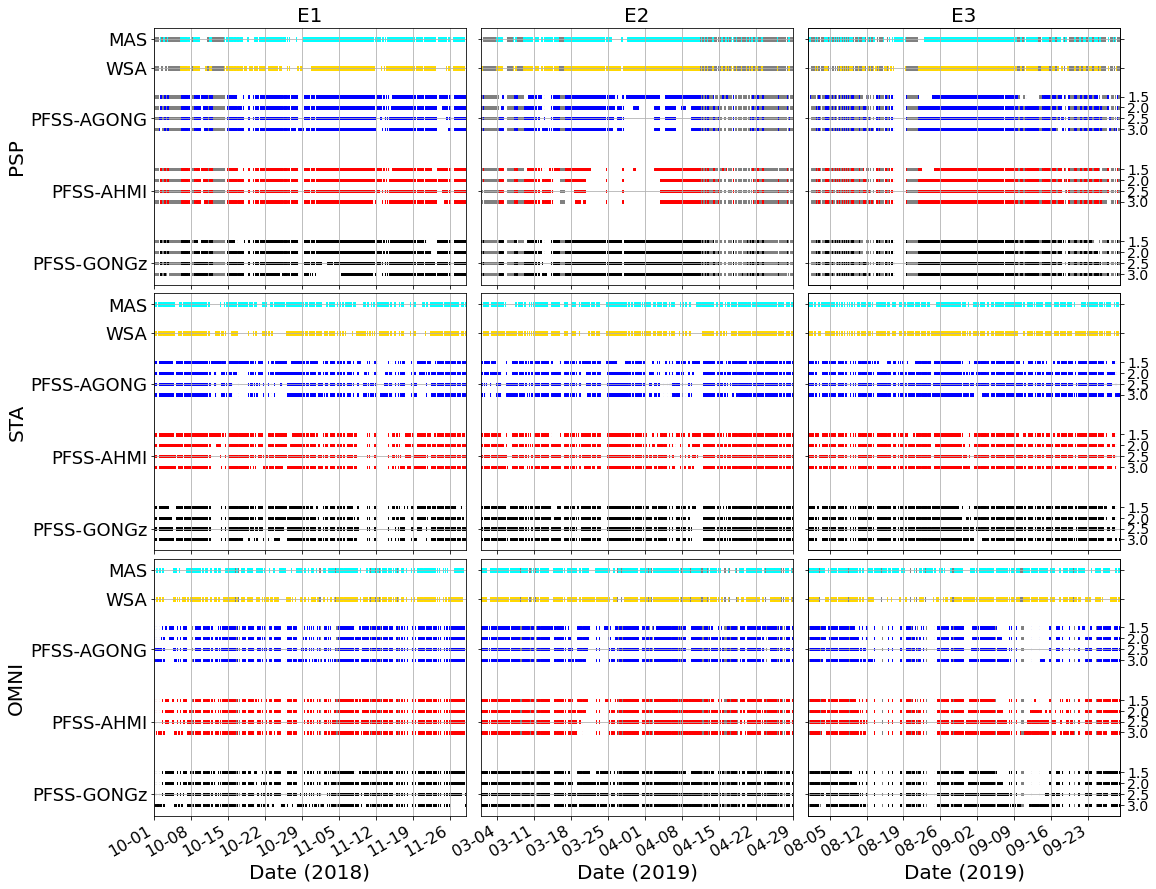}         
   \caption{\textit{In situ} magnetic polarity and prediction comparison as a function of time for each encounter. Metric 3 produces a single score for each encounter as described in the main text, Section \ref{subsec:metric3} so cannot be represented as a timeseries. This figure depicts how the predicted and measured polarity agreed or disagreed as a function of time and therefore produced a high or low metric score. Columns from left to right show encounters 1,2 and 3 respectively, while each row show results from different spacecraft sources: PSP, OMNI and STEREO A respectively. In each panel, results for the different models are shown, following the color scheme from figure \ref{fig:results2-vs_model}. For the PFSS models, there is a different line of data shown for each source surface height value, ordered from low to high. At an hourly cadence, if the model and measured polarity agree then that point is filled in with the respective color, if the model and measured polarity disagree, that point is left blank, and if \textit{in situ} measurements are missing for a given time, they are colored grey. Therefore, gaps in the lines indicate intervals of disagreement between model and measurement. The larger the fraction of the whole time series which is blank, the lower the resulting metric score reported in figures \ref{fig:results1-vs_rss} and \ref{fig:results2-vs_model}.
   }
   \label{fig:results4-3_vs_time}
\end{figure*}

We conclude our exposition of the metric score results by examining the \textit{in situ} polarity metric in detail. As mentioned above, this metric is formed by integrating  measured and modeled polarity timeseries together and therefore only yield one score per encounter. However, we can examine qualitatively how these scores are formed by examining the timeseries of agreement and disagreement for each model and encounter combination.

This information is encapsulated in figure \ref{fig:results4-3_vs_time}. Each panel shows model agreement and disagreement for a given encounter, and given heliospheric spacecraft. For times when the model and data agree, the time series has a colored marker (with color indicating differing models) and is blank if they disagree. For times when there is not \textit{in situ} data available, the markers are colored grey. Therefore, times when the measured and modeled polarities disagree (or equivalently, when the positions of predicted HCS crossings are misaligned with the data) may be identified as gaps in the colored lines on the plot. For the PFSS models, lines for different source surface heights are laid out consecutively (from low to high source surface height). The growth or decay of the white gaps on these consecutive lines indicates if a given HCS crossing improves or worsens with changing source surface height. 

Overall, it is clear that most spacecraft data and model HCS combinations agree very well for the vast majority of timestamps, which accounts for the high metric scores. Notable exceptions are with PSP in Encounter 2 and OMNI data for Encounter 3 for which there are much larger gaps in agreement indicating severe misalignment between data and model HCS.   

For the PFSS models, we see that typically the disagreements (blank spaces) occur at the same time \textit{for a given spacecraft in all models} and also usually follow the same qualitative trend as a function of source surface height. This indicates that small misalignments between HCS warps and the empirical data correspond to the same heliographic locations independent of the input magnetogram. The MAS and WSA models sometimes have gaps aligned in time with the PFSS models (for example PSP Encounter 1 10/29, top left panel), but often do not, suggesting the HCS output of these models can be qualitatively different to the PFSS models and sometimes in better agreement with spacecraft data. A very clear example of this is in the top middle panel (E2, PSP) where several large outages in agreement for PFSS models at all source surface heights occur while the MAS and WSA models both correctly predict the polarity of PSP throughout.    

The timeseries are usually very different between different spacecraft. The gaps in agreement can occur at different times for a given model. This is not surprising since the spacecraft are at different solar longitudes and crossing the HCS at different times. However, there are also examples where one spacecraft gives much better agreement than another. Encounter 3 is the clearest case of this where results for the OMNI data set show large intervals of disagreement, while PSP and STEREO A have much more continuous agreement (this may be verified by referring to the ordering of the spacecraft scores in the bottom right panel of figure \ref{fig:results2-vs_model}). Another difference between different spacecraft is that for PSP the agreement overall is much more continuous, while for the 1AU datasets it is more intermittent with agreement switching on and off from one hour to another. This is caused by the relatively larger fluctuation amplitudes and Parker spiral angle in the $B_R$ data at 1AU (OMNI and STEREO A), meaning that the measured timeseries include fluctuations in polarity which do not correspond to current sheet crossings.

This concludes our exposition of the metric scoring results in this work. We next discuss the implications of these results, both on the specific time intervals, models and magnetograms studied here, as well as what they show about this framework for evaluating coronal models.

\section{Discussion and Conclusions}
\label{sec:discussion}

In this work, we have defined a model-agnostic framework to evaluate and rank coronal models based on their magnetic field topology using three independent sources of observational data. A given coronal model needs only to provide 2D maps of (a) open magnetic field at the photosphere and (b) magnetic polarity at the model outer boundary. Models are given a metric score based on 1) how well they represent coronal holes as observed in EUV data for the relevant Carrington rotation, 2) how well they predict streamer belt topology observed by white light coronagraph observations, 3) how well they predict heliospheric current sheet crossing timing measured \textit{in situ} by various heliospheric spacecraft.  

We applied our framework to a dataset of coronal models for the first three encounters of Parker Solar Probe using Potential Field Source Surface (PFSS) models for a range of source surface heights and input magnetograms, as well as concurrent Wang-Sheeley-Arge (WSA) models and the Magnetohydrodynamic Algorithm outside a Sphere (MAS) models.  We carried out three investigations. The first focused on PFSS models only and explored the metric scores as a function of source surface height and choice of magnetogram inner boundary condition. The second compared the PFSS model scores to WSA and MAS scores, evaluating the impact of more physics in the model. The third examined the time evolution of the metric scores in a 60 day interval for each PSP encounter period to look at the stability of the results.

For the PFSS-only investigations, we found : \newline

(A1) \textit{Quality of representation of coronal holes at the photosphere is a function of source surface height:} 

Values above $2.5$ $R_\odot$ lead to observed coronal holes in EUV not being captured by the model (low precision) regardless of input boundary condition. Values too low can also lead to opening fields in the model which aren't observed in EUV (low recall), although for models driven by the GONGz data product, the best scores occurred for the lowest source surface height. This indicates that for the same source surface height and observation date, ADAPT maps produce more open flux, as was previously observed in \citep{Badman2020}. Examining appendix Figure \ref{fig:AppendixA_Metric1_E1}, the optimal score for a given magnetogram occurs when the model and observed boundaries of the polar coronal holes are close together. For ADAPT maps this occurs near 2.0$R_\odot$ and below this the polar coronal holes are overexpanded, while for GONGz, the best match is found at 1.5$R_\odot$. This suggests the coronal hole metric defined here is sensitive to the polar flux treatment in the input magnetogram. The best F-scores were typically around 60\%, requiring both reasonable polar and low latitude coronal hole representation. \newline

(A2) \textit{Representation of streamer belt topology is also strongly dependent on the source surface height of PFSS models, but the dependence is in tension with the coronal hole representation:}

For all magnetogram choices and time intervals, the representation improves monotonically with increasing source surface heights, although with diminishing returns above 2.5$R_\odot$. This variation is simply explained by how the warps in the HCS produced by PFSS models are smoothed out with increasing source surface heights. White light Carrington maps of the streamer belt depict a typical low solar activity HCS with warps at most 40 degrees above or below the solar equator. \newline

(A3) \textit{Prediction of HCS crossing times is shown to be only weakly dependent on source surface height.}

This is interesting since we are testing the same model observable as in the streamer belt metric (HCS location), but against a different data source (\textit{in situ} magnetic polarity). While metric (2) is investigating amplitudes of warps in the HCS, this metric is investigating the longitudinal position of these warps along the heliographic equator. Therefore we show that warps in the HCS predicted by PFSS are typically in the correct heliographic position, regardless of their amplitude, with the correct spacecraft polarity indicated more than 60\% of the time and as much as 90\% of the time for certain orbits and spacecraft. Is is worth noting that this metric (specifically as applied to 1AU spacecraft) is very similar to the basis for the canonical source surface height of $2.5R_\odot$ established by \citet{Hoeksema1983,Hoeksema1984}. We see several examples where this canonical height does give the best polarity score but it is not universally true and the score difference between 2.0 or 2.5 $R_\odot$ is typically very small.\newline

\textit{(A4) A ``global optimum source surface height'' is not in general possible with PFSS models of the solar corona.}

Combining our inferences, we see that while it is certainly possible to optimize coronal hole representation, or representation of warps in the HCS, these cannot be achieved simultaneously. This impasse is indicative of the bare-bones physical content of PFSS and provides a simple explanation for why different authors who have focused on different physical coronal signatures \citep[e.g. ][]{Hoeksema1983,Lee2011,Arden2014} have arrived at different optimum values for the source surface. Nevertheless, we can also see why PFSS modeling is useful and widely used: with generally accepted source surface height parameters (2.0-2.5 Rs), such models can produce good (if not optimum) metric scores in all 3 metrics. Simultaneous good representation of coronal holes and HCS crossing timing allows for accurate solar wind source mapping and comparison to \textit{in situ} data.

By extending our investigation from PFSS to include other models with additional physics, we found the following : \newline

(B1) \textit{Inclusion of full MHD physics via the MAS model \textit{globally} improves model scores for metrics (1) and (2), showing MHD predictions of coronal hole locations and general shape of the HCS are more in agreement with observational evidence than PFSS models with any configuration of input parameters.} 

This shows, unsurprisingly, that the physics present in MHD does produce a more accurate representation of the magnetic structure of the corona at both the inner and outer boundary. At least for encounter 1, there is a marked asymmetry in the dominant warp in the HCS in the MHD model coming from differential flow velocity \citep{Riley2019}, providing an example of the physics present in such models that are neglected in potential models. \newline

(B2) \textit{Comparison of WSA results to PFSS shows global improvement in metric 2 (streamer belt representation) but similar results in the other metrics.} 

This is not surprising since what WSA adds to PFSS models is a Schatten Current Sheet (SCS) layer, which slightly intersects the PFSS model domain near the source surface. It is thus not expected to drastically change which coronal loops open up to the solar wind, but it does have a strong effect on the HCS beyond the source surface, flattening it. As mentioned earlier, coronagraph observations of the streamer belt show the HCS is flatter or less warped than PFSS models for $R_{ss} \leq 2.0R_{S}$. Therefore WSA offers a solution to the lack of a global optimum source surface height in PFSS models, while maintaining easy computation: the source surface height could be tuned to optimize coronal hole representation and the WSA SCS layer would then take care of sufficiently flattening the HCS, allowing both inner and outer model boundary conditions to better match observations. \newline

(B3) \textit{For E1 and E3 metric 3 showed little dependence on the model chosen, and some evidence for MHD improvement for E2 where there was solar activity.}

This indicates that under solar minimum conditions, the location of warps in the HCS in heliographic longitude are fairly accurately reproduced by PFSS models and do not really change with an addition of more physics. The exception to this for our study was encounter 2 where there was significant solar activity and far side evolution in the magnetograms, and therefore less solar-minimum like conditions. Here, the score for specifically Parker Solar Probe was significantly improved in the MHD model, but was also dramatically better in GONGz vs the ADAPT PFSS models. This suggests that this difference may be due to innaccurate modeling of far side evolution in the ADAPT magnetograms as well as a lack of accurate model physics. \newline

Overall, then we conclude that our framework for coronal model evaluation successfully allows comparison between models with very different constituents, and moreover shows expected differences between models with different physical content. 

Lastly, the time evolution of the metric scores (where available) showed : \newline

(C1) \textit{Metric scores do evolve in time with a different profile seen for each time interval studied, but profiles are typically very similar between models using the same magnetogram.}

The score ordering between models is therefore preserved as a function of time. This suggests the time evolution is due to new information being introduced into the magnetogram data or remote observations as the Sun rotates which affects the model coronal field in qualitatively the same way between models. This has the implication that far side evolution of the photospheric field, and therefore divergence of the model inner boundary condition from reality, is a fundamental limitation to coronal modeling no matter the physics that is introduced. \newline

(C2) \textit{Far side flux emergence can lead to sudden changes in metric scores.}

The importance of far side evolution is particularly emphasized by PSP encounter 2 (middle column, figure \ref{fig:results3-12_vs_time}), where step changes in metric scores are observed (around 2019-03-20 and 2019-04-08). For this specific encounter, there was significant solar activity \citep[][]{Pulupa2020} which evolved and changed the global coronal field while out of view from Earth-based magnetograms, mostly concentrated in two key active regions \citep[AR 12737 and AR12378, ][]{Harra2021,Cattell2021}. This meant a strong photospheric field concentration rotated on disk twice during this time interval and at this time produced sudden and significant changes in magnetograms. This produced simultaneous sharp changes both in the accuracy of representation of coronal holes, and in representation of the global streamer belt topology (both reductions in accuracy as implied by the metrics defined in this work). This change was also observed as a loss of accuracy in metric 3 (middle column figure \ref{fig:results4-3_vs_time}), most dramatically in PSP data but also to some extent in 1AU data, indicating severe distortion to the model HCS topology by this flux emergence.\newline

(C3) \textit{The agreement between PSP \textit{in situ} data and the model predictions was typically more continuous than that of STEREO A and OMNI}. 

This was illustrated in figure \ref{fig:results4-3_vs_time} and demonstrates a simple advantage of sampling the solar wind closer to the Sun for the application of evaluating coronal models : the amplitude of magnetic vector fluctuations to the base value is reduced and this means a more reliable and continuous measurement of which side of the HCS a given spacecraft is located. \newline

In summary, the focus of this work has been defining a framework for evaluating coronal models using several independent physical observables. A key ingredient was the requirement for models to produce a consistent output (2D maps of the heliospheric current sheet). This means the authors of future coronal models can easily compare to older models and verify if they produce improved results, and if they only improve certain aspects of the coronal field representation of if the improvement is global across different metrics. We demonstrated this with application to a small subset of models centered around the first three Parker Solar Probe perihelia and were able to compare PFSS, WSA and MHD models, investigate the source surface height parameter to PFSS models, and observe the impact of far side evolution and flux emergence on coronal model accuracy. 

Having specified this evaluation method, there are many examples of future work that can be tackled. We close with listing a few examples.

This work has entirely focused on 3 time intervals around solar minimum (2018-2019) driven by our use of novel PSP data. As the solar cycle waxes, the behavior of all three metrics on different models will be very interesting to study. For example, it is expected that the accuracy of PFSS models should decay due to the prevalence of non-potential structures in the active Sun, while MHD like models should be more robust.

There is also a wide parameter space of models that should be investigated. The MAS MHD model studied in this work is at the upper end of computational effort and complexity and there are intermediate codes such as PLUTO \citep{Mignone2007}, MS-FLUKSS \citep{Pogorelov2013} or AWSoM \citep{vanderHolst2014} that could be cross-compared. Even within the MAS framework, several energetics models exist \citep{Riley2021e} that could be evaluated for impacts on the coronal magnetic structure, thereby allowing these metrics to test different theories of coronal heating and solar wind acceleration. Such investigations could be quite powerful in directly addressing fundamental coronal physics questions.

Further studies of magnetogram-magnetogram comparisons are also clearly motivated by this work. Although we have gone some way towards studying this here through our comparisons of ADAPT-GONG, ADAPT-HMI and GONGz maps applied to the same models, there remain a large number of other magnetogram sources with different polar coronal hole assumptions or different surface flux transport models and different empirical corrections to correct for line of sight or magnetograph saturation effects \citep[e.g. ][]{Wang2022}. Additionally, we did not touch here on the question of magnetogram resolution. All maps used here had 1 x 1 degree binning and just by visual inspection of e.g. Figure \ref{fig:AppendixA_Metric1_E1} we see this is more than sufficient to pick out detailed structure of both polar and equatorial coronal holes. Nevertheless, both sharper (e.g. HMI)  and coarser (e.g. Wilcox Solar Observatory) resolution maps are available. \citet{Caplan2021} showed how even with potential models,  high resolution magnetograms can be used to produce fine-scale coronal structure which is lost at lower resolution. This framework could be used to evaluate the benefit of having increased resolution in terms of matching observational data.

Finally, although not a specific study, a clear further direction for this work is in the creation of an open source tool or web application to perform metric evaluations. This tool could be used by any coronal modeler by submitting the coronal hole and neutral line maps in the format prescribed here and are then presented with the resulting metric scores and supporting plots. While such an integrated software tool is beyond the scope of the current project, we presented in boxes \ref{alg:metric1},\ref{alg:metric2} and \ref{alg:metric3} descriptions of the algorithm to produce respective metric score. We also supply the base observational data products to perform the comparisons for the time intervals studied in this work as a Zenodo repository \citep{Zenodo2022} in an effort to allow other modelers to try evaluating this framework on their own models. As well as the specific observational data products, python scripts are included for each metric as examples for loading and plotting the different file types.


\acknowledgments
      We acknowledge support from ISSI for the team 463, entitled 'Exploring The Solar Wind In Regions Closer Than Ever Observed Before'. Parker Solar Probe was designed, built, and is now operated by the Johns Hopkins Applied Physics Laboratory as part of NASA's Living with a Star (LWS) program (contract NNN06AA01C). Support from the LWS management and technical team has played a critical role in the success of the Parker Solar Probe mission. The FIELDS and SWEAP experiments on Parker Solar Probe spacecraft were designed and developed under NASA contract NNN06AA01C. The authors acknowledge the extraordinary contributions of the Parker Solar Probe mission operations and spacecraft engineering teams at the Johns Hopkins University Applied Physics Laboratory.  This work utilizes data obtained by the Global Oscillation Network Group (GONG) Program, managed by the National Solar Observatory, which is operated by AURA, Inc. under a cooperative agreement with the National Science Foundation. The data were acquired by instruments operated by the Big Bear Solar Observatory, High Altitude Observatory, Learmonth Solar Observatory, Udaipur Solar Observatory, Instituto de Astrofísica de Canarias, and Cerro Tololo Interamerican Observatory. This work utilizes data produced collaboratively between Air Force Research Laboratory (AFRL) \& the National Solar Observatory (NSO). The ADAPT model development is supported by AFRL. The input data utilized by ADAPT is obtained by NSO/NISP (NSO Integrated Synoptic Program). This research made use of HelioPy, a community-developed Python package for space physics \cite{Stansby2019b}. This research has made use of SunPy v2.0.1 \citep{Mumford2019}, an open-source and free community-developed solar data analysis Python package \citep{Sunpy2015}. The work of DHB and HPW was performed under contract to the Naval Research Laboratory and was funded by the NASA Hinode program. S.D.B. acknowledges the support of the Leverhulme Trust Visiting Professorship program. STB. was supported by NASA Mary W. Jackson Headquarters under the NASA Earth and Space Science Fellowship Program Grant 80NSSC18K1201. The work of AR and NP was funded by the ERC SLOW{\_}\,SOURCE project (SLOW{\_}\,SOURCE - DLV-819189). We acknowledge use of NASA/GSFC’s Space Physics Data Facility’s OMNIWeb service and OMNI data.

%





\clearpage
\appendix

\section{Full Model-Observation Comparisons}
\label{sec:appendixA}

This appendix contains summary plots of the observation-model comparisons for all metrics considered in the work and are animated as a function of time for each encounter in the online version.

\subsection{Metric 1 : Coronal Hole - Open Field Comparison}
\label{subsec:AppA-metric1}

\begin{figure*}[ht!]
   \centering
   \includegraphics[width=0.9\textwidth]{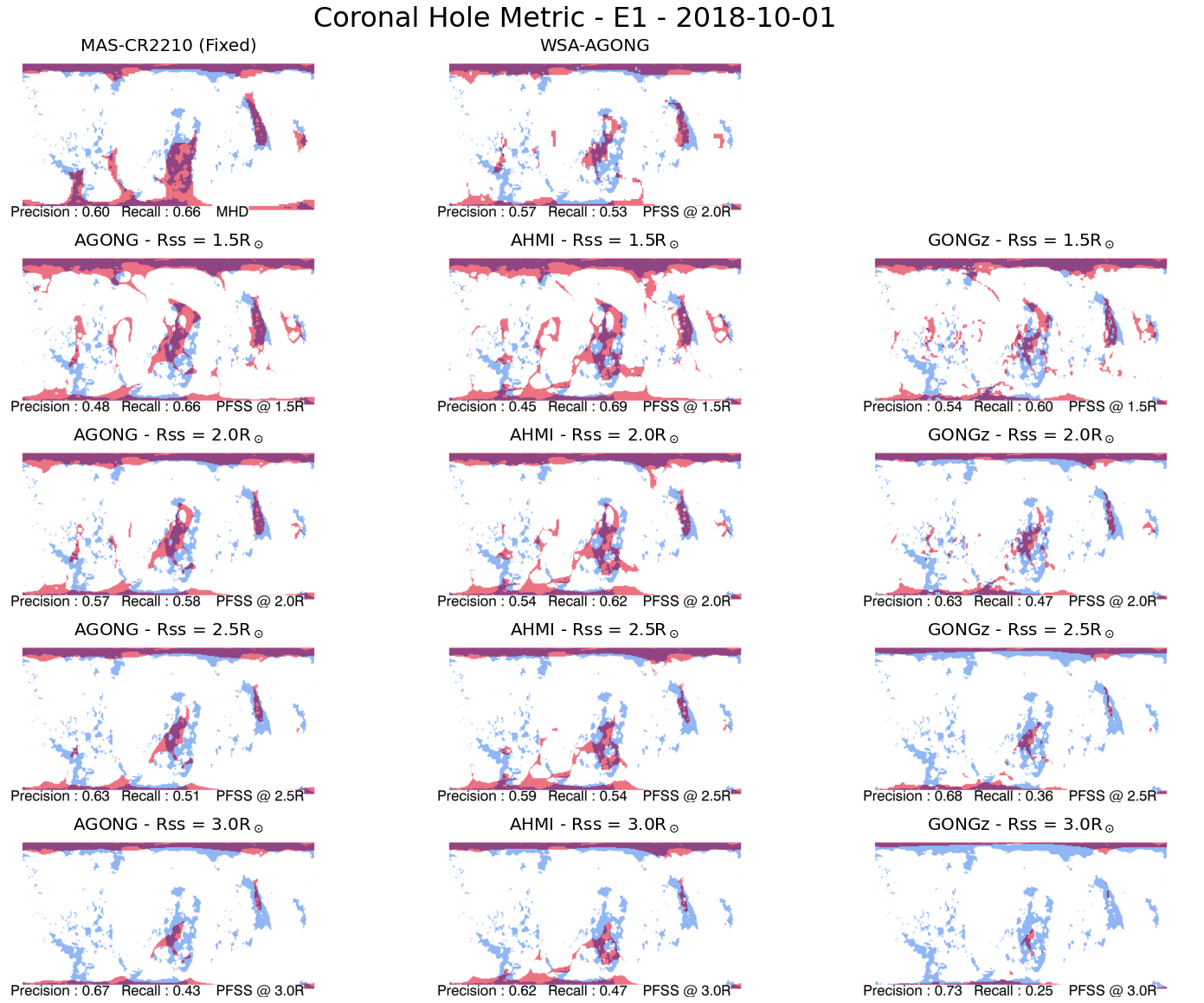} 
   \caption{Metric 1 (Coronal Hole) model-observation comparison for the first PSP encounter. The panels depict the binary classification scheme used to define our coronal hole metric, as detailed in section \ref{subsec:metric1} and illustrated in the bottom panel of figure \ref{fig:metric1-illustration}. Blue shading indicates the coronal hole area determined by applying the EZSEG algorithm to EUV synoptic maps, red shading indicates the open field area determined from the particular coronal model run. Where the model and observations agree, the pixels are shaded purple. The titles of each panel indicate the model and model parameters. The top row shows MAS and WSA results, while the bottom four rows show PFSS results with the columns differentiating the model and the rows differentiating the source surface height. Inset text in each panel records the associated precision and recall scores. In the online material, this figure is animated as a function of time for the encounter from October-1-2018 to November-30-2018 at a daily cadence. Although the background EUV map is fixed, the model coronal hole distributions evolve except for the MAS MHD results (top-left panel) which are based on a single  magnetogram for this encounter.}
   \label{fig:AppendixA_Metric1_E1}%
\end{figure*}

\begin{figure*}[ht!]
   \centering
   \includegraphics[width=0.9\textwidth]{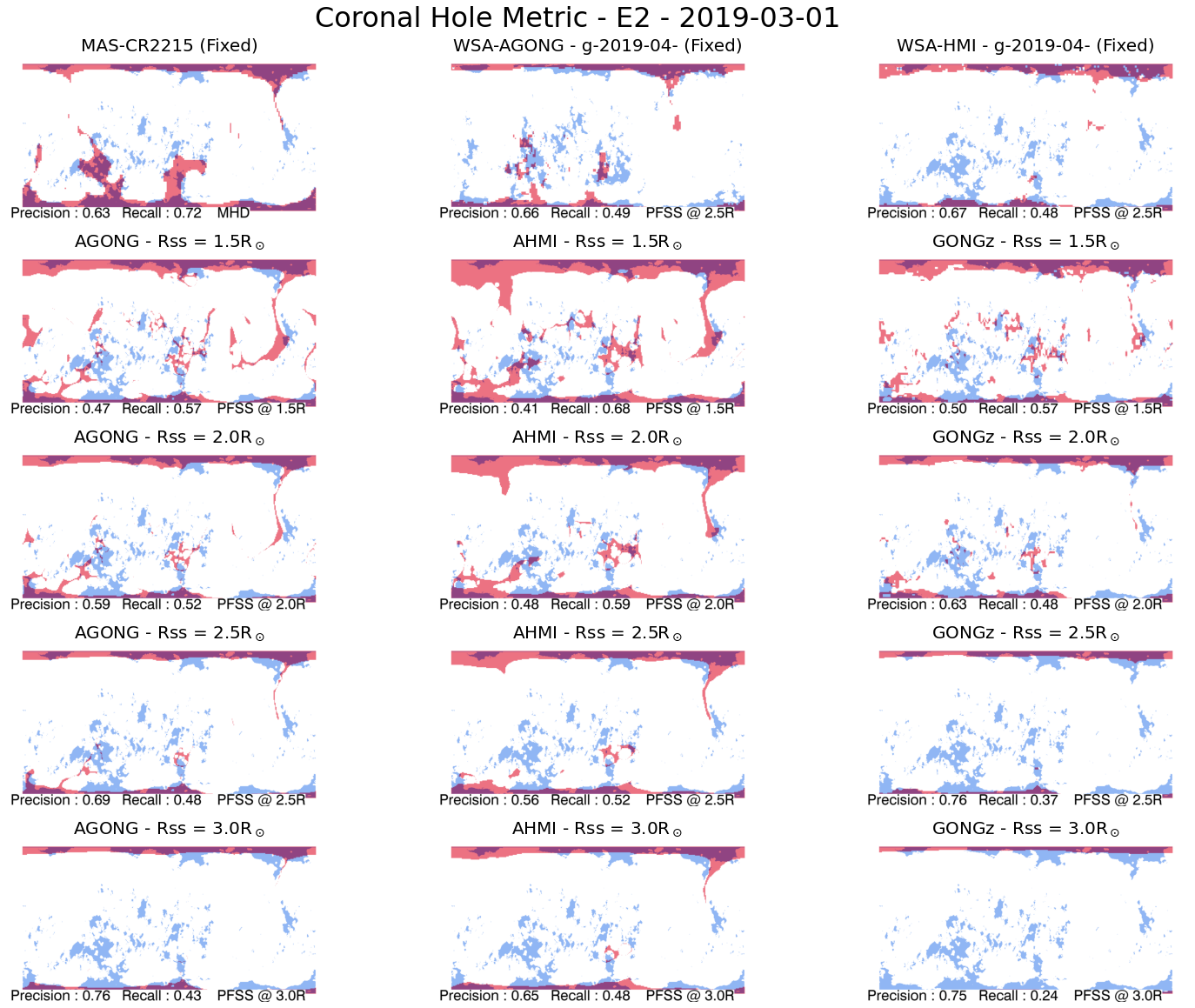}         
   \caption{Metric 1 (Coronal Hole) model-observation comparison for the second PSP encounter. The panels are organized as in figure \ref{fig:AppendixA_Metric1_E1}. The online animation shows daily evolution from 2019-03-01 to 2019-04-30 except for the MAS and WSA models (top row) which are based on a single magnetogram for this encounter.}
   \label{fig:AppendixA_Metric1_E2}%
\end{figure*}

\begin{figure*}[ht!]
   \centering
   \includegraphics[width=0.9\textwidth]{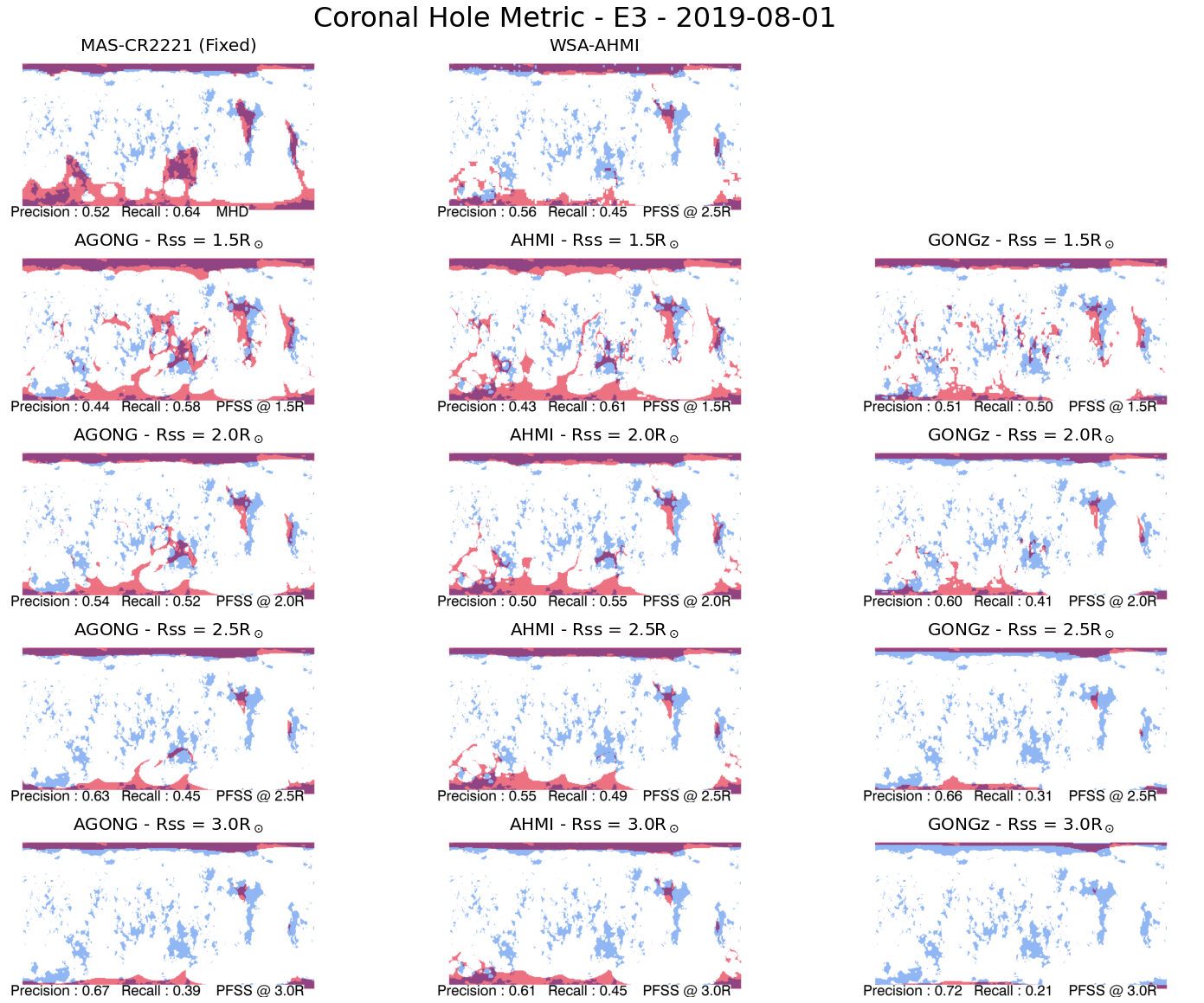}         
   \caption{Metric 1 (Coronal Hole) model-observation comparison for the third PSP encounter. The panels are organized as in figure \ref{fig:AppendixA_Metric1_E1}. The online animation shows daily evolution from 2019-08-01 to 2019-09-30 except for the MAS model (top-left panel) which are based on a single magnetogram for this encounter.}
   \label{fig:AppendixA_Metric1_E3}%
\end{figure*}

Here, in figures \ref{fig:AppendixA_Metric1_E1}-\ref{fig:AppendixA_Metric1_E3} we show schematics of the binary classification from EZSEG and the coronal model open field maps detailed in section \ref{subsec:metric1}. The panels are organized to differentiate the models and (for PFSS models), the source surface height. Each panel has the precision and recall shown as a text inset. As a reminder, the quantity quoted in the main text (section \ref{sec:results}), the `F-score', is the harmonic mean of these two quantities. These figures are animated over each 60 day interval in the online version of this manuscript. The still images shown here are the first frame of the movie from October-01-2018, March-01-2019 and August-01-2019 for encounters 1-3 respectively.

\subsection{Metric 2 : Neutral Line - Streamer Belt Comparison}
\label{subsec:AppA-metric2}

\begin{figure*}[ht!]
   \centering
   \includegraphics[width=0.9\textwidth]{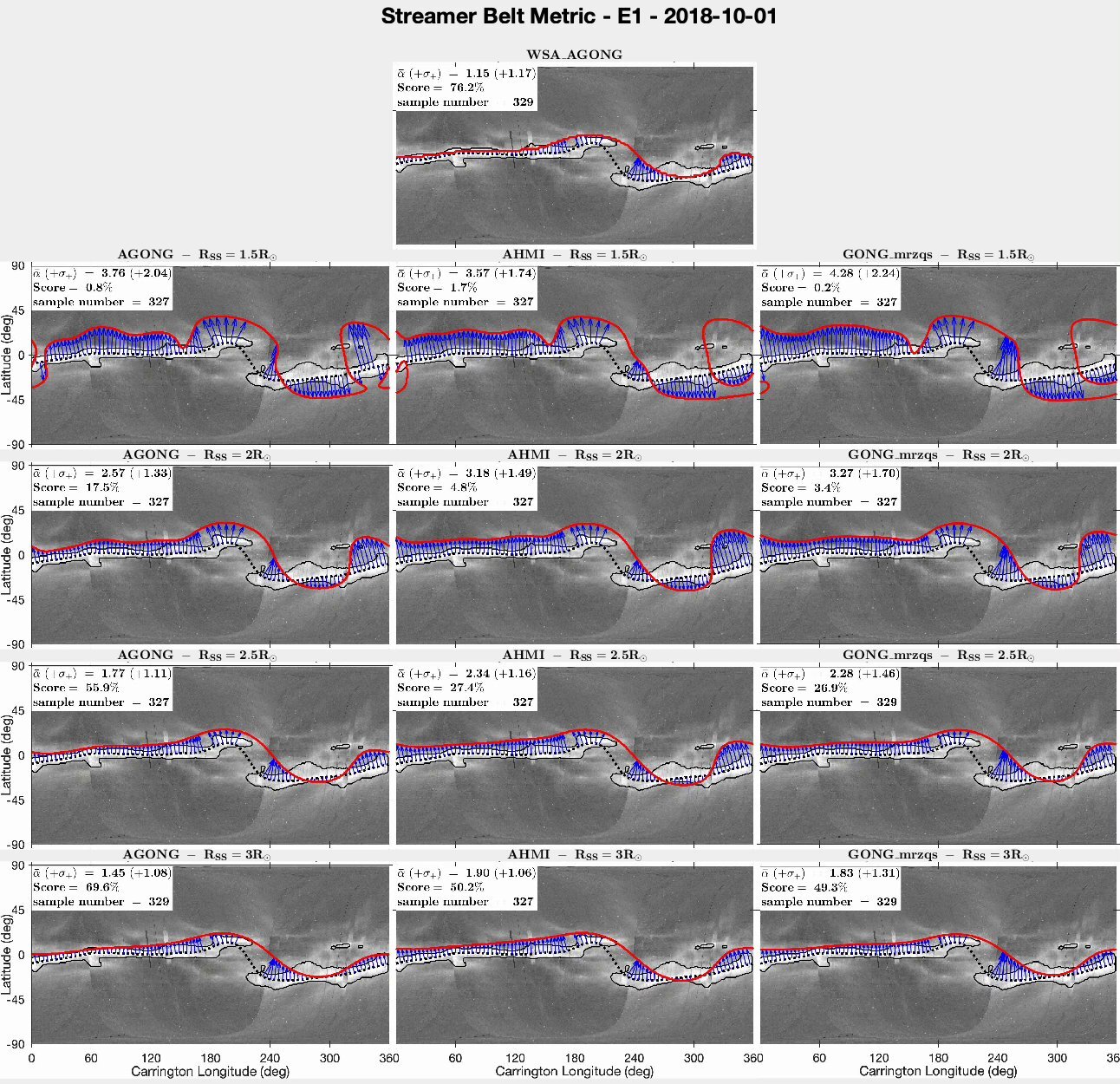}         
   \caption{Metric 2 (Streamer belt) model-observation comparison for the first PSP encounter. The panels show the comparison between the extracted streamer belt line and the model HCS, as detailed in section \ref{subsec:metric2} and illustrated in the bottom panel of figure \ref{fig:metric2-illustration}. The streamer belt score as well as the intermediate $\bar{\alpha}$ and $\sigma_+$ quantities are given in the top-left corner of each panel. In the online version of this manuscript, this figure is animated at a daily cadence from October-01-2018 to November-30-2018, showing how the model neutral lines and background Carrington white light (WL) maps evolve as a function of time (except the MAS model, top left panel, for which only the WL map evolves in this encounter).}
   \label{fig:AppendixA_Metric2_E1}%
\end{figure*}

\begin{figure*}[ht!] 
  \centering
  \includegraphics[width=0.9\textwidth]{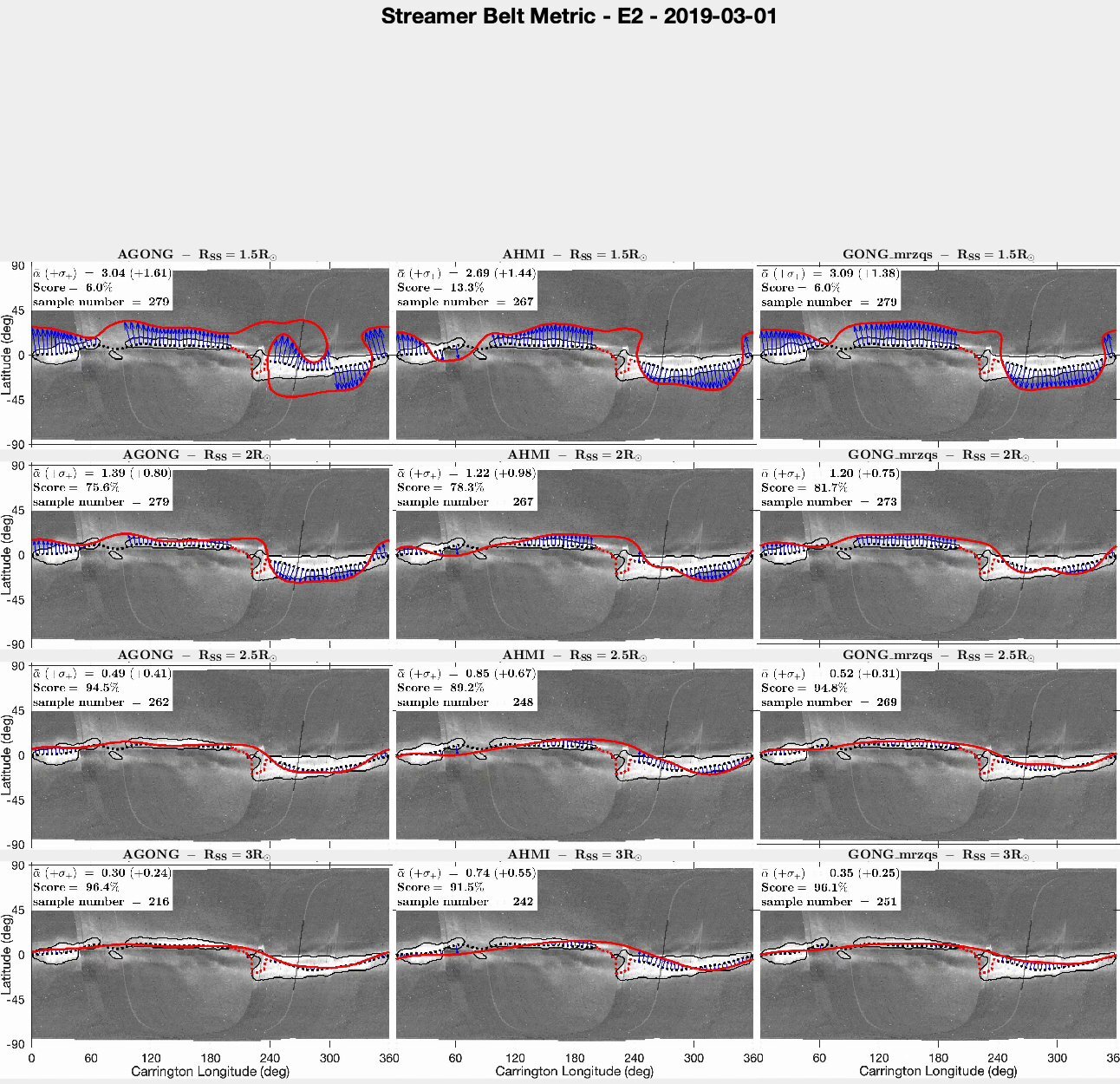}         
  \caption{Metric 2 (Streamer belt) model-observation comparison for the second PSP encounter. The panels are organized as in figure \ref{fig:AppendixA_Metric2_E1}. In the online version of this manuscript, this figure is animated at a daily cadence from March-01-2019 to April-30-2019, except for the MAS and WSA models, (top row), for which only the WL map evolves.}
  \label{fig:AppendixA_Metric2_E2}%
\end{figure*}

\begin{figure*}[ht!]
  \centering
  \includegraphics[width=0.9\textwidth]{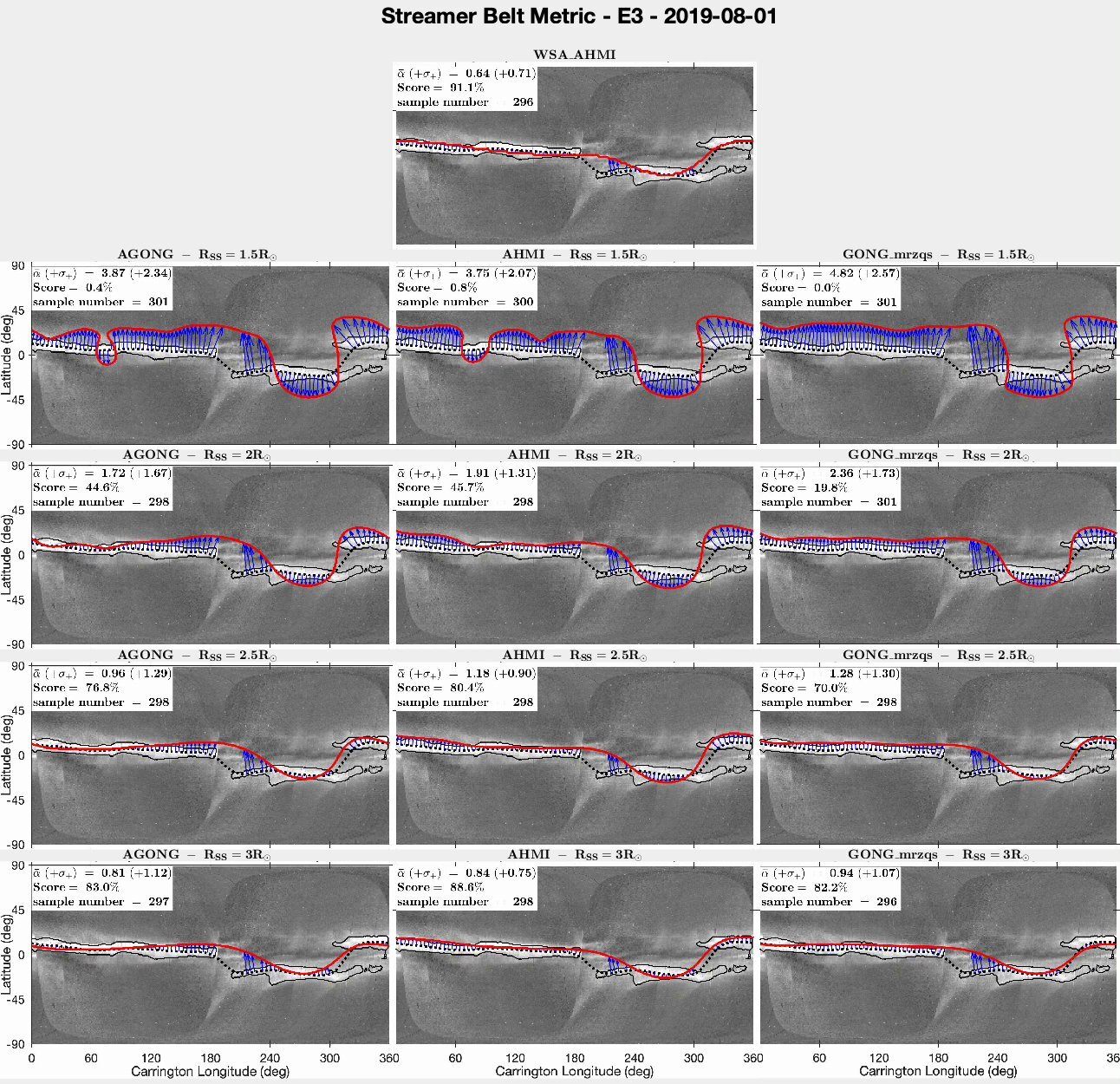}         
  \caption{Metric 2 (Streamer belt) model-observation comparison for the third PSP perihelion. The panels are organized as in figure \ref{fig:AppendixA_Metric2_E1}. In the online version of this manuscript, this figure is animated at a daily cadence from August-01-2019 to September-30-2019, except for the MAS models (top-right panel) for which only the WL map evolves.}
  \label{fig:AppendixA_Metric2_E3}%
\end{figure*}

Following the format of section \ref{subsec:AppA-metric1}, we next show schematically the comparison between the modeled and measured streamer belt topology. Again we show a multi-panel figure for each encounter (figures \ref{fig:AppendixA_Metric2_E1}-\ref{fig:AppendixA_Metric2_E3}) with panels organized by model and source surface height. Each panel follows the same format as the bottom panel of figure \ref{fig:metric3-illustration}: the background image, black contour and dashed centroid curve show the white light (WL) carrington map and extracted streamer belt (SMB line), while the red curve shows the model HCS and the blue lines show the distances between the curves which contribute to the metric score, as detailed in section \ref{subsec:metric2} and \citet{Poirier2021}. In the online version of this manuscript, these figures are movies for each  encounters over the 60 day intervals (in which both elements evolve with time for the 60 day interval around perihelion, except where noted in the figure captions). The static versions of these figures show the first frame of each movie from dates October-01-2018, March-01-2019 and August-01-2019 for encounters 1-3 respectively.

\subsection{Metric 3 : \textit{In sit}u polarity - HCS Comparison}
\label{subsec:AppA-metric3}

\begin{figure*}[ht!]
   \centering
   \includegraphics[width=0.9\textwidth]{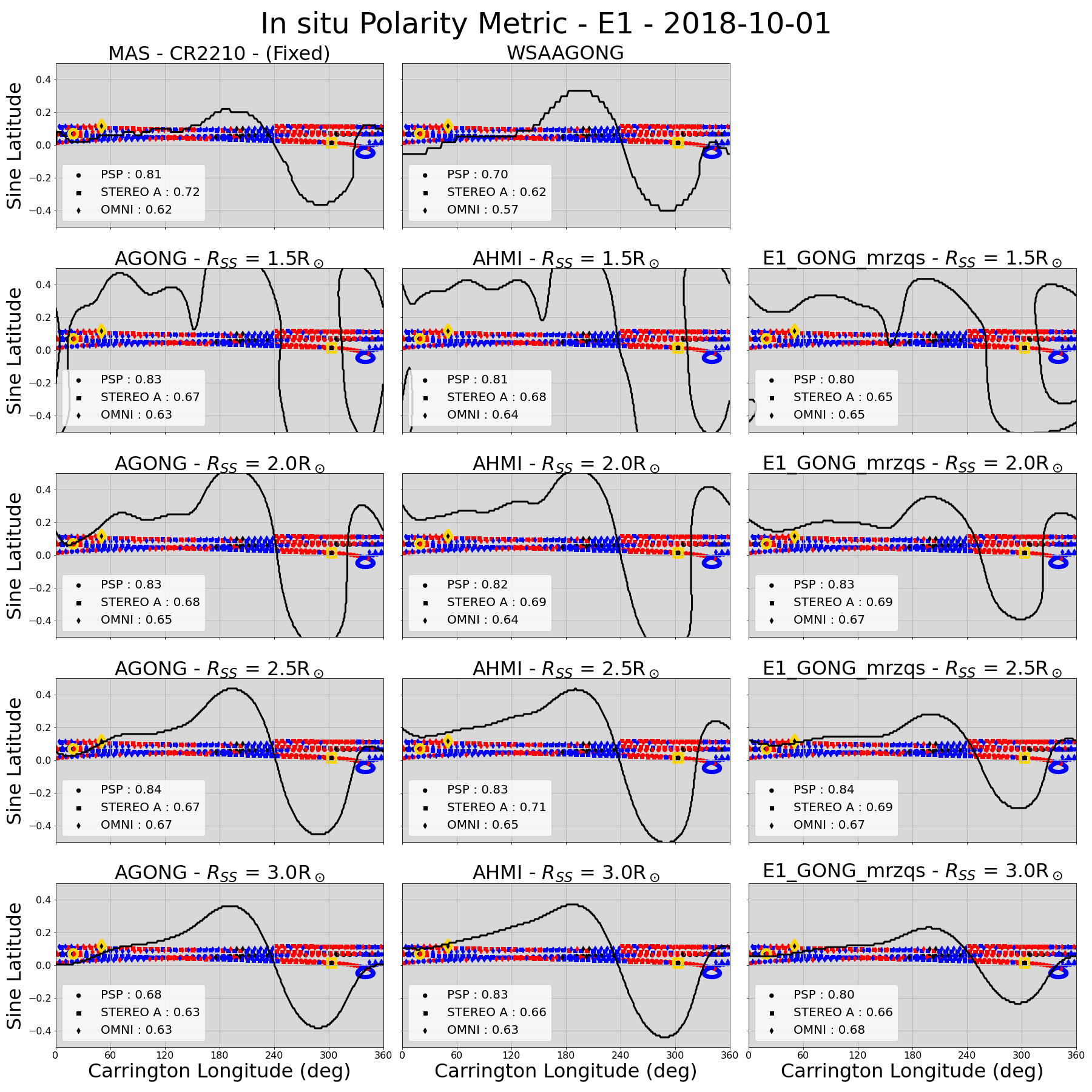} 
   \caption{\textit{In situ} polarity model-observation comparison for models for the first PSP encounter. The panels schematically show the extent to which the \textit{in situ} measured polarity (blue/red markers) corresponds to the modeled heliospheric current sheet (black contour), as detailed in section \ref{subsec:metric1} and illustrated in the middle panel of figure \ref{fig:metric3-illustration}. Model-data agreement occurs where the \textit{in situ} data is taken northwards of a HCS contour \textit{and} is red (positive polarity) or where the data is taken southwards of the HCS contour \textit{and} is blue (negative polarity). The panel titles indicate the model parameters. The top row shows MAS and WSA results, while the bottom four rows show PFSS results with the columns differentiating the model and the rows differentiating the source surface height. In the online edition, this figure is animated for the 60 day interval hith each frame advancing in time one day at a time from October-01-2018 to November-30-2018. As time evolves, the model HCSs evolve for each model (except for the MAS model, top left panel which only has a single model run for E1). Simultaneously, the spacecraft (yellow symbols) move in the solar corotating reference frame. The static version of the figure is the first frame for the video, showing the metrics for the date October-01-2018.}
   \label{fig:AppendixA_Metric3_E1}%
\end{figure*}

\begin{figure*}[ht!]
   \centering
   \includegraphics[width=0.9\textwidth]{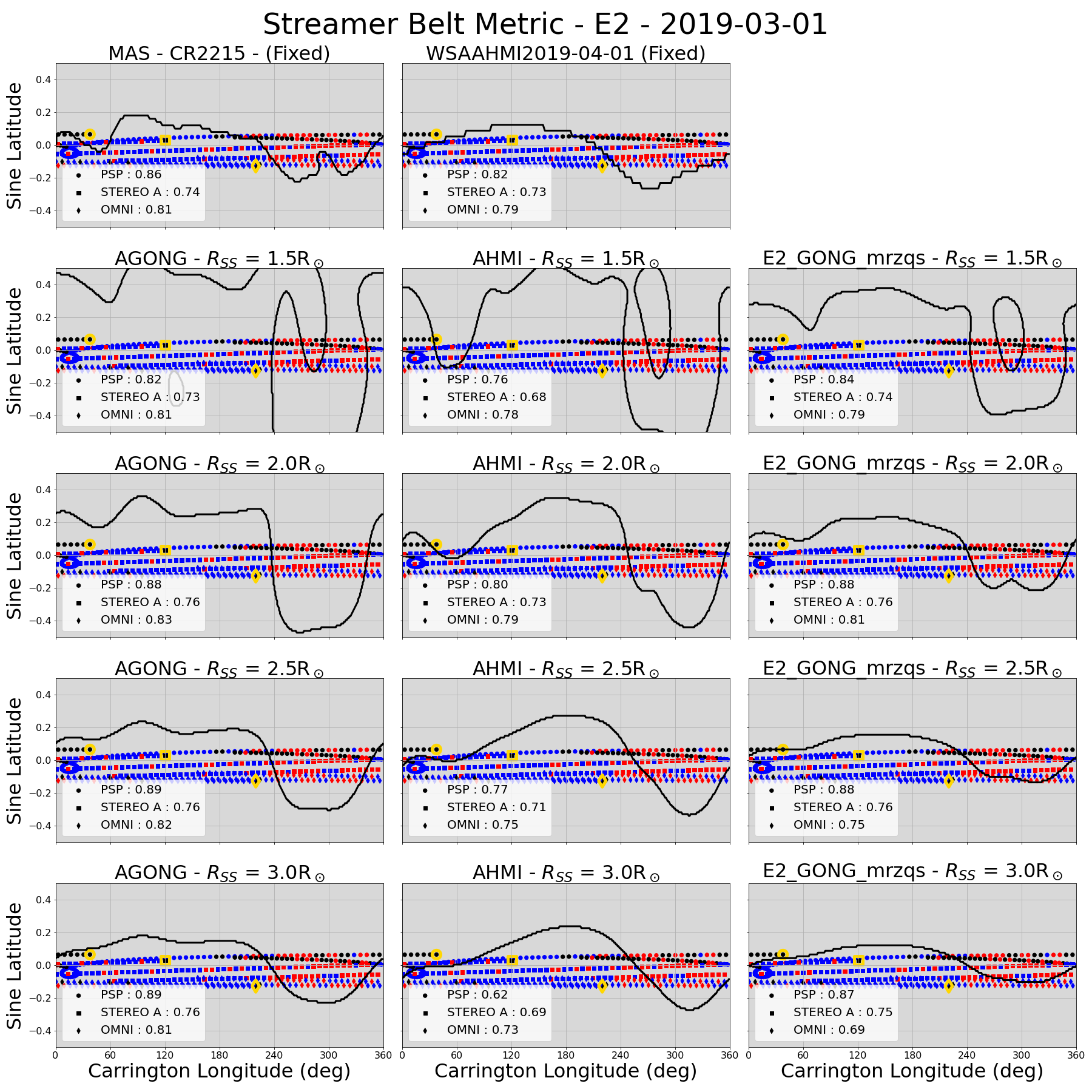} 
   \caption{\textit{In situ} polarity model-observation comparison for models for the second PSP encounter. The panels are organized as in figure \ref{fig:AppendixA_Metric3_E1}. In the online edition of this manuscript, this figure is animated for the 60 day interval for encounter 2 from March-01-2019 to April-30-2019 showing relative motion of the model HCS and spacecraft positions (projected to the source surface). Here the model HCS from MAS and WSA (top row) are frozen in time.. The static version of the figure is the first frame for the video, showing the metrics for the date March-01-2019.}
   \label{fig:AppendixA_Metric3_E2}%
\end{figure*}

\begin{figure*}[ht!]
   \centering
   \includegraphics[width=0.9\textwidth]{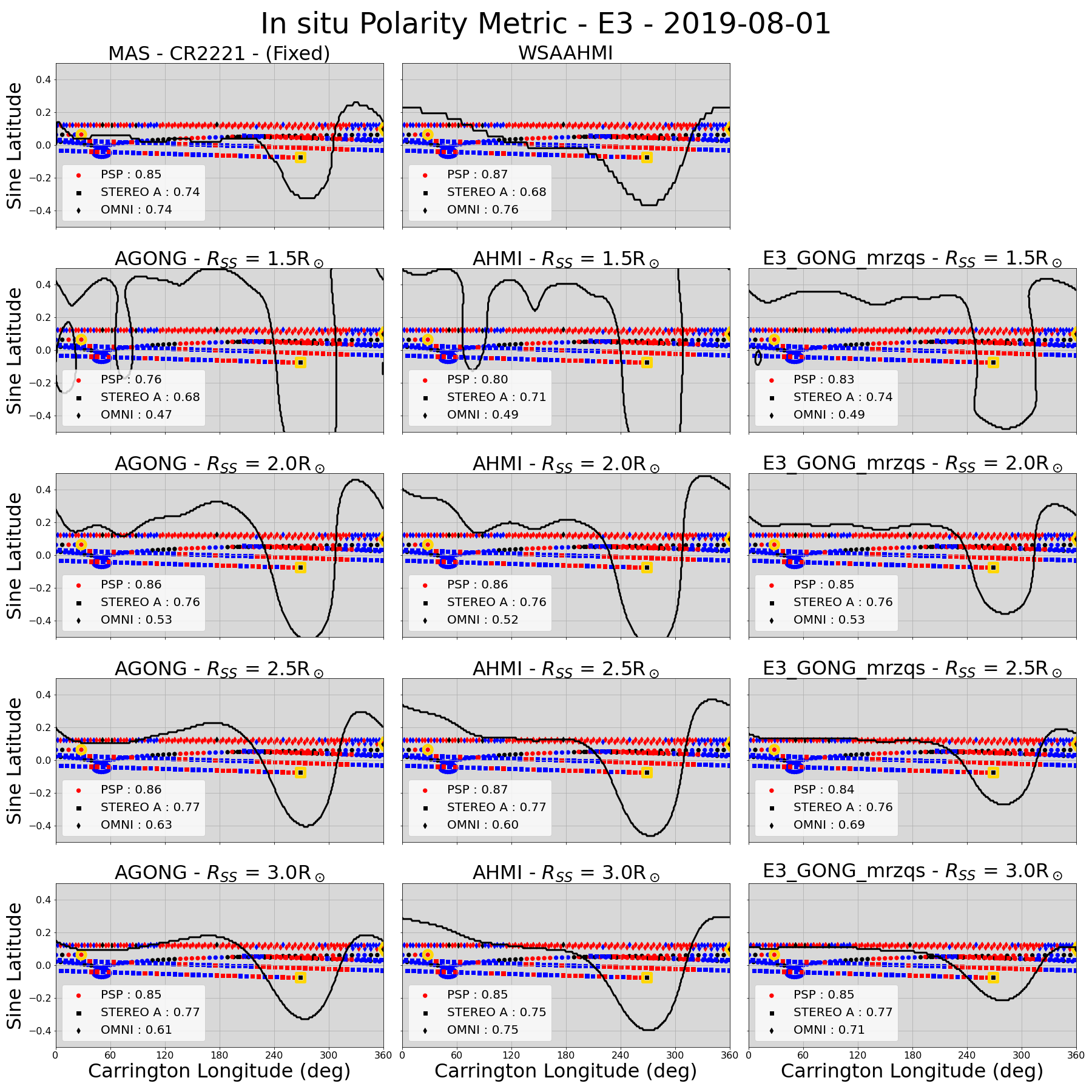} 
   \caption{\textit{In situ} polarity model-observation comparison for models for the third PSP encounter. The panels are organized as in figure \ref{fig:AppendixA_Metric3_E1}. In the online edition of this manuscript, this figure is animated for the 60 day interval for encounter 3 from August-01-2019 to September-30-2019, showing relative motion of the model HCS and spacecraft positions (projected to the source surface). Here the model HCS from MAS (top-left panel) is frozen in time. The static version of the figure is the first frame for the video, showing the metrics for the date August-01-2019.}
   \label{fig:AppendixA_Metric3_E3}%
\end{figure*}

Finally, in figures \ref{fig:AppendixA_Metric3_E1}-\ref{fig:AppendixA_Metric3_E3} we summarize model-observation comparisons for the third metric studied (\textit{in situ} polarity). These figures are corollaries to figure \ref{fig:results4-3_vs_time} which already displayed in a compact way all model/data comparisons for this metric as function of time. Here we plot the measured data in the format of the middle panel of the illustrative figure \ref{fig:metric3-illustration}, showing the measured polarity timeseries data plotted as the spacecraft trajectory ballistically mapped across the model outer boundary. This is to allow the reader to see visually how the change in the model's HCS topology affects agreement with the \textit{in situ} data.

As in sections \ref{subsec:AppA-metric1} and \ref{subsec:AppA-metric2}, here we have an individual multi-panel plot for each encounter for all model and model input parameters studied in this work. In the online versionof this manuscript, these figures are animated over the relevant 60 day interval for each encounter, showing the movement of the HCS as the models evolve (except where noted in the figure captions) as the spacecraft whose \textit{in situ} data we use move in the heliographic frame. The still images shown here are the first frame of theese movies, snapshotting  October-01-2018, March-01-2019 and August-01-2019 for encounters 1-3 respectively.


\bibliography{issi_corona}
\bibliographystyle{aasjournal}



\end{document}